\documentclass[prd,aps,floats,onecolumn,twoside,preprintnumbers,
superscriptaddress,floatfix,nofootinbib,showpacs]{revtex4}
\usepackage{graphicx,pstricks,rotating,amsmath,hyperref}
\usepackage{slashed}

\begin{document} 

\title{\center{Photon self-interaction on deformed spacetime}}

\author{Raul Horvat}
\affiliation{Institute Rudjer Bo\v{s}kovi\'{c}, Bijeni\v{c}ka 54 10000 Zagreb, Croatia}
\email{raul.horvat@irb.hr}
\author{Josip Trampeti\'{c}}
\affiliation{Max-Planck-Institut f\"ur Physik, (Werner-Heisenberg-Institut), F\"ohringer Ring 6, D-80805 M\"unchen, Germany}
\email{trampeti@mppmu.mpg.de,josipt@rex.irb.hr;
On leave of absence from Institute Rudjer Bo\v{s}kovi\'{c}, Zagreb, Croatia}
\author{Jiangyang You}
\affiliation{Institute Rudjer Bo\v{s}kovi\'{c}, Division of Theoretical Physics, Bijeni\v{c}ka 54 10000 Zagreb, Croatia.}
\email{youjiangyang@gmail.com}

\newcommand{\tr}{\hbox{tr}} 


\begin{abstract}
A novel nonlocal four-photon interaction on the deformed spacetime is derived and studied
in the three selected models (I, II, III). The first two models (I, II) are obtained
via two distinct second-order $\theta$-exact Seiberg-Witten maps of the
noncommutative U(1) gauge field strength on Moyal space. The third one (III),
inspired by the manifestly gauge invariant structures emerging in the first
two, due to the model generality has been constructed with a different set of freedom parameters. The physical relevancy of all models  is analyzed by evaluating the four-photon-tadpole diagram, which, when combined with the bubble graph, enables us to fully
consider all contributions to the one-loop photon polarization tensor. For an arbitrary noncommutative matrix $\theta^{\mu\nu}$, the full quadratic IR divergence cancellation in the one-loop photon two-point function is obtained with particular combinations of Seiberg-Witten map/gauge-symmetry freedom parameters in models I and II. Finally, our model III enables complete elimination of all pathological terms in the photon polarization tensor at one-loop order if a special value for the $\theta$ matrix is chosen.
\end{abstract}

 \pacs{11.10.Nx, 11.15.-q, 12.10.-g}



\maketitle

\section{Introduction}

The application of the $\theta$-exact Seiberg-Witten map (SW)~\cite{Seiberg:1999vs}
expansions~\cite{Jurco:2000fb,Madore:2000en,Jurco:2000ja,Mehen:2000vs,Szabo:2001kg,Jurco:2001my,Jurco:2001rq,Barnich:2002pb,Barnich:2003wq,Banerjee:2003vc,Banerjee:2004rs,Martin:2012aw,Martin:2015nna} represents the current state-of-the-art in the field of the noncommutative (NC) gauge field theories \cite{Schupp:2008fs,Horvat:2010km,Horvat:2011bs,Horvat:2011qg,Aschieri:2012in,Aschieri:2014xka,Horvat:2013rga} and the associated particle-physics phenomenology
~\cite{
Horvat:2010sr,Horvat:2011iv,Horvat:2012vn,Wang:2012ye,Wang:2013iga}. A summation over all orders in the antisymmetric tensor $\theta^{\mu\nu}$ at tree level is automatically achieved via this approach, leading to various interesting results. More importantly, the $\theta$-exact approach allows an access to nonlocal effects within the perturbative approach, most pronouncedly the quadratic UV/IR mixing~\cite{Minwalla:1999px,Hayakawa:1999yt,Hayakawa:1999zf,Matusis:2000jf} in the two-point functions at one loop~\cite{Schupp:2008fs,Horvat:2013rga}.

A particular topic related to the SW map approach to the noncommutative gauge theories is the implementation of the gauge freedoms into the (inter)action. This step is generally regarded as favorable as far as the control over various divergences in the perturbative quantum corrections is concerned. The original employment was via a $\theta$-iterative construction~\cite{Bichl:2001cq}. Later on a $\theta$-exact substitute was suggested~\cite{Trampetic:2014dea} and generalized to a second-order expansion afterwards~\cite{Trampetic:2015zma}.

While after the first order, it is natural to consider the second-order SW map
(either in the perturbative  or in the $\theta$-exact approach) in the perturbative
quantum field theory, in the past the second order has been  much less investigated~
\cite{Alboteanu:2007bp,Schupp:2008fs}. The reason was obviously technical: The second-order SW map solution is inherently much more complicated than the first-order one and requires consequently more effort to obtain explicitly the gauge invariant action and to implement the gauge freedom. Recently the model building works based on the $\theta^2$ order SW map of non-Abelian gauge fields  have received  more attention~
\cite{Aschieri:2012in,Dimitrijevic:2012pk,Aschieri:2014xka,Dimitrijevic:2014iwa} when
the order-$\theta$  correction vanishes, yet studies on quantum corrections are still
absent. Going from the $\theta$-expansion to $\theta$ exact, the second-order SW
map adds its unique additional difficulties. Two expansion solutions sharing the
identical first order do exist~\cite{Mehen:2000vs,Martin:2012aw}. Each solution involves its own type of 3-products  ($\star_3$ and $\star_{3'}$ as in~\cite{Trampetic:2015zma}), and while the leading order of these two solutions with respect to $\theta$ can be shown to be connected by gauge transformation, the full solutions are not~\cite{Trampetic:2015zma}.

The gauge freedom structure at second order in the $\theta$-exact approach is considerably more complicated than the first order, and also more complicated than its $\theta^2$-order counterpart. Besides the existence of two solutions, the field strength expansion from each solution contains distinct gauge freedom structures~\cite{Trampetic:2015zma}. Analyzing the $\theta^2$ order indicates that more gauge freedom structures will show up after performing the integration-by-part in the action. Performing the same
procedure $\theta$-exactly meets the difficulty from the noncommutativity and/or
nonassociativity of the generalized star products. This issue is much less pressing at the $e^2$ order since the $\star_2$ product satisfies the so-called 3-cyclicity~\cite{Mylonas:2013jha}
\begin{equation}
\int f(x)\big(g(x)\star h(x)\big)=\int h(x)\big(g(x)\star f(x)\big),\;\;\int f(x)\big(g(x)\star_2 h(x)\big)=\int h(x)\big(g(x)\star_2 f(x)\big).
\label{3-cy}
\end{equation}
However, there is no 4-cyclicity in general, --($\star_2$ and $\star_{3'}$ products have some cyclicity left, while $\star_3$ has practically no cyclicity left.) In this paper we will show that the proper substitute to integration-by-part in the $\theta$-exact computation on Moyal space is to  Fourier-transforming the whole computation to the momentum space and achieve the explicit gauge-invariant action simultaneously with the verification of the Ward identity of the vertex.

Following the preliminary work in~\cite{Trampetic:2015zma}, we present in this paper, for
the first time, the full SW-map based/inspired nonlocally deformed four-photon
couplings on Moyal space, with all reasonable gauge freedom parameters included, and their quantum correction induced by such coupling via the one-loop four-photon-tadpole diagram. We use two distinct second-order $\theta$-exact SW map expansions for the gauge field strength (which we call model (I) and (II), as dubbed recently in~\cite{Trampetic:2015zma}) with all the freedom/ambiguity/deformation parameters included. From these field strengths we then derive the corresponding actions and show that they can be expressed explicitly in terms of the commutative $\rm U(1)$ field strength $f_{\mu\nu}=\partial_\mu a_\nu-\partial_\nu a_\mu$ by working out the integration-by-part procedure required. Our final form for both SW map based actions further indicates that the $\theta$-exact freedom parameters we suggested before~\cite{Trampetic:2014dea} would not deplete all the $\theta$-iterative possibilities in~\cite{Bichl:2001cq}. The two additional gauge freedom/ambiguity/deformation parameters are thus introduced by hand in turn.

We then determine the four-photon coupling vertices with all possible gauge freedom/ambiguity/deformation parameters included from models I and II and write down the one-loop four-photon-tadpole integrals for these models. We find that various momentum factors in the second-order SW map expansion reduce either to unity  or to the common nonlocal factor $\sin^2\frac{p\theta k}{2}/\big(\frac{p\theta k}{2}\big)^2$ in the tadpole, the same as found in the three-photon-bubble diagram studied previously~\cite{Horvat:2013rga}.

It is long known that the massless tadpole integrals all vanish at the integration dimensions $D\ge 4$ under dimensional regularization~\cite{Leibbrandt:1975dj}. However this result is modified by the nonlocal factors~\cite{Hayakawa:1999zf,Horvat:2011qg}. In order to precisely verify this effect we evaluate the tadpole integral using two different methods: The first method simply introduces a pair of identical numerators and denominators to turn the tadpole into the bubble integral then evaluate the tadpole using the protocol established for the bubble diagram, as we did in~\cite{Horvat:2013rga}. The second method generalizes the $n$-nested zero regulator~\cite{Leibbrandt:1975dj} for the commutative tadpole diagram.  We compute the four-photon-tadpole contributions to the photon two-point function as a function of unspecified number of the integration dimensions $D$. Then we specify gauge field theory dimension d by taking the limits $D\to d$. Next we especially discuss the $d=4$ case. In the end, we find that both approaches reveal the same {\em purely} quadratic IR divergent result in the $D\to 4-\epsilon$ limit, verifying the soundness of our computation.

The UV/IR mixing phenomenon, reflecting the inherent nonlocality of the full theory and arising from the high-momentum region of integration in the Feynman integrals, shows up as a IR divergence when the spatial extension of the NC string of size $|\theta p|$ gets reduced to a point. A related anomaly is a nonanalytical behavior in the NC parameter $\theta$, when the limit $\theta \rightarrow 0$ is undertaken. In the NC gauge theory, however, a quadratic IR divergence coexists with the logarithmic divergence which matches the UV behavior. Our study indicates that the quadratic IR divergence is clearly connected with tadpole integrals. Therefore the gauge invariant four photon interaction we found may serve as counterterms to cancel quadratic IR divergence. For this purpose the tadpole induced quadratic IR divergence is summed together with the corresponding contribution from the photon bubble diagram~\cite{Horvat:2013rga}. We find that while it is indeed possible to do so in both models I and II, the procedure requires fixing the first-order gauge freedom parameter~\cite{Trampetic:2014dea} $\kappa=1$. Subsequently a third possible action (III), inspired by the structures of the first two, is introduced. It involves a SW-map inspired gauge invariance deformation in a more general way.

In this action each manifestly gauge invariant four-photon coupling term starting at $\theta^2$ order is assigned an independent freedom parameter, which is shown to be sufficient to cancel any quadratic IR divergences from the bubble diagram for the two typical $\theta$ values. (The UV divergences still require separated fine-tuning.)

This paper is structured as follows:
In the first two sections we describe two $\theta$-exact Seiberg-Witten map models up to the $e^3$ order and construct the four-photon self-interaction, as a model definition.  Section III is devoted to the computation, presentation, and discussion of the $\theta$-exact tadpole in $D$ and four dimensions, while in the Sec. IV we analyze a sum of the bubble and the tadpole diagram to show elimination of IR divergences for arbitrary $\theta$-matrix. In the Sec. V we introduce a generalized model of the deformed four-photon interaction and show the elimination of all divergences from the bubble plus the tadpole diagram for the special choice of the $\theta$-matrix and freedom parameters. Sections VI and VII present the discussion and conclusion, respectively. In this article the capital letters denote noncommutative objects, while the small letters denote the commutative ones.

\section{A model definition of the three- and the four-photon self-interactions}

\subsection{Definitions and construction of the actions}

As usual we consider the formal $\rm U_{\star}(1)$ NC gauge theory action
\begin{equation}
S=-\frac{1}{4e^2}\int F_{\mu\nu}\left(e a_\mu,\theta^{\mu\nu};\kappa,\eta,...\right)
\star F^{\mu\nu}\left(e a_\mu,\theta^{\mu\nu};\kappa,\eta,...\right),
\label{2.00}
\end{equation}
where the formal NC gauge field strength
$F_{\mu\nu}\left(e a_\mu,\theta^{\mu\nu},\kappa,\eta,...\right)$
is regarded as a composite operator built-up using the commutative gauge field operator $a_\mu$ and the NC parameter $\theta^{\mu\nu}$ via the SW map procedure. The set of parameters $(\kappa,\eta,...)$ represents in principle the SW map/gauge-invariance freedoms. The commutative coupling constant $e$ is attached to the commutative gauge field operator $a_\mu$ due to the charge quantization issue~\cite{Horvat:2011qn}. As a bonus feature it also serves as the ordering parameter for the $\theta$-exact SW map expansion, i.e.
\begin{equation}
F_{\mu\nu}=e f_{\mu\nu}+F^{e^2}_{\mu\nu}+F^{e^3}_{\mu\nu}+\mathcal O\left(e^4\right)
\label{2.0}.
\end{equation}
To $e^2$ order the gauge field strength SW map $F^{e^2}_{\mu\nu}$ expansion is fairly universal
\cite{Horvat:2013rga,Trampetic:2014dea,Trampetic:2015zma}
\begin{equation}
F^{e^2}_{\mu\nu}=e^2\theta^{ij}
\Big(\kappa f_{\mu i}\star_{2}f_{\nu j}-a_i\star_2\partial_j
f_{\mu\nu}\Big).
\label{2.1}
\end{equation}
Note that $\kappa$ deformation in the settings of this paper (\ref{2.1}) and in a recent works  \cite{Trampetic:2014dea,Trampetic:2015zma}, corresponds to the ${\kappa^{-1}_g}$ for the previous  $\kappa_g$ deformation of \cite{Horvat:2013rga} \footnote{See also a discussion after Eq (26) in \cite{Trampetic:2014dea}. We first substitute $\kappa_g=\kappa^{-1}$, then extract out from all terms in the action a factor
$\kappa^{-2}$. That factor is than absorbed as an overall rescaling of the field redefinition~\cite{Trampetic:2014dea}. Further on we name this paper settings the $\kappa$-settings.}. In this paper the $\kappa$-deformation is dubbed the $\kappa$-settings, where we have the following triple-photon action:
\begin{equation}
S^{e}=-\frac{e}{2}\int\theta^{ij}f^{\mu\nu}\left(\kappa f_{\mu i}\star_2 f_{\nu j}-\frac{1}{4}f_{ij}\star_2f_{\mu\nu}\right)\,,
\label{3photon}
\end{equation}
responsible to the contribution to the photon polarization tensor arising from the photon bubble diagram \cite{{Horvat:2013rga}}.

The (profound) structure of the $\theta$-exact SW map of a $\rm U(1)$ gauge theory is summarized in \cite{Trampetic:2015zma}, where two distinct gauge field SW maps were found and analyzed up to the $e^3\sim a_\mu^4$ order. Expanding \eqref{2.00} up to order $a_\mu^4$ with (\ref{2.0}) gives the following general form for the four-photon interaction
\begin{equation}
S^{e^2}=-\frac{1}{4e^2}\int\,F^{e^2}_{\mu\nu}F^{ e^2\mu\nu}+2ef^{\mu\nu}F_{\mu\nu}^{e^3},
\label{2.2}
\end{equation}
where the following two distinct solutions for the $e^3$ order gauge field strength have been found and given explicitly in \cite{Trampetic:2015zma}. The first solution is resolved from SW differential equation  \cite{Trampetic:2015zma},
\begin{equation}
\begin{split}
F^{e^3}_{\mu\nu_{\rm (I)}}&(x)_{\kappa,\kappa_1,\kappa_2}=
\frac{e^3}{2}\theta^{ij}\theta^{kl}\bigg[\kappa_1\left(\left[f_{\mu k}f_{\nu i} f_{l j}\right]_{\star_{3'}}+\left[f_{\nu l}f_{\mu i}f_{kj}\right]_{\star_{3'}}\right)-\kappa a_i\star_2\partial_j\left(f_{\mu k}\star_2 f_{\nu l}\right)
\\&-\kappa_2\left(\left[f_{\nu l}a_i\partial_j f_{\mu k}\right]_{\star_{3'}}
+\left[f_{\mu k}a_i\partial_j f_{\nu l}\right]_{\star_{3'}}+\left[a_k\partial_l\left(f_{\mu i}f_{\nu j}\right)\right]_{\star_{3'}}-2a_i\star_2\partial_j\left(f_{\mu k}\star_2 f_{\nu l}\right)\right)
\\&+\left[a_i\partial_j a_k \partial_l f_{\mu\nu}\right]_{\star_{3'}}
+\left[\partial_l f_{\mu\nu}a_i\partial_j a_k\right]_{\star_{3'}}+\left[a_k a_i \partial_l\partial_j f_{\mu\nu}\right]_{\star_{3'}}
-\frac{1}{2}\Big(\left[a_i\partial_k a_j\partial_l f_{\mu\nu}\right]_{\star_{3'}}
+\left[\partial_l f_{\mu\nu}a_i\partial_k a_j\right]_{\star_{3'}}\Big)\bigg]\,,
\label{2.3}
\end{split}
\end{equation}
while the second one is obtained by inverting the known solution for the inverted SW map  \cite{Mehen:2000vs} in \cite{Trampetic:2015zma},
\begin{equation}
\begin{split}
F^{e^3}_{\mu\nu_{\rm (II)}}&(x)_{\kappa,\kappa'_1,\kappa'_2}
=e^3\theta^{ij}\theta^{kl} \Big[\kappa'_1\left(f_{\mu i}\star_2\left(f_{jk}\star_2 f_{l\nu}\right)+f_{l\nu}\star_2\left(f_{jk}\star_2 f_{\mu i}\right)-\left[f_{\mu i}f_{jk}f_{l\nu}\right]_{\star_3}\right)
\\&-\kappa'_2\big((a_i\star_2\partial_j f_{\mu k})\star_2 f_{\nu l}+(a_i\star_2\partial_j f_{\nu l})\star_2 f_{\mu k}-[a_i\partial_j (f_{\mu k}f_{\nu l})]_{\star_3}\big)
\\&- \kappa a_i\star_2\partial_j\left(f_{\mu k}\star_2 f_{\nu l}\right)+(a_i\star_2\partial_j a_k)\star_2\partial_l f_{\mu\nu}
\\&+a_i\star_2(\partial_j a_k\star_2\partial_l f_{\mu\nu})+a_i\star_2(a_k\star_2\partial_j\partial_l f_{\mu\nu})-[a_i\partial_j a_k\partial_l f_{\mu\nu}]_{\star_{3}}
\\&-\frac{1}{2}\Big(a_i\star_2(\partial_k a_j\star_2\partial_l f_{\mu\nu})+(a_i\star_2\partial_k a_j)\star_2\partial_l f_{\mu\nu}-[a_i\partial_k a_j\partial_l f_{\mu\nu}]_{\star_3}
+[a_ia_k\partial_j\partial_l f_{\mu\nu}]_{\star_3}\Big)\Big].
\label{2.4}
\end{split}
\end{equation}
Definitions of generalized star products and the momentum dependent
functions $f_{\star_2}\left(p,q\right)$, $f_{\star_3}\left(p,q,k\right)$
and $f_{\star_{3'}}\left(p,q,k\right)$ are given in the Appendix A.

In both solutions we have included the following freedom parameters:
$\kappa$ from $F^{e^2}_{\mu\nu}$, while in $F^{e^3}_{\mu\nu}$ we have $(\kappa,\kappa_{1,2})$ for model I and $(\kappa,\kappa'_{1,2})$ for model II, respectively. From those field strengths we have found the following two actions at the $a_\mu^4$ order,
\begin{equation}
\begin{split}
S^{e^2}_{\rm (I)}=&-\frac{e^2}{4}\theta^{ij}\theta^{kl}\int\,\kappa^2(f_{\mu i}\star_2 f_{\nu j})(f^\mu_{\;\,\; k}\star_2 f^\nu_{\;\,\; l})-\kappa(f_{ij}\star_2 f_{\mu\nu})(f^\mu_{\;\,\; k}\star_2 f^\nu_{\;\,\; l})
+2\kappa_1f^{\mu\nu}[f_{\mu i}f_{\nu k}f_{jl}]_{\star_{3'}}
\\&
+2\kappa_2 f^{\mu\nu}(a_i\star_2\partial_j(f_{\mu k}\star_2 f_{\nu l})-[f_{\mu k} a_i\partial_j f_{\nu l}]_{\star_{3'}}-[a_i f_{\mu k}\partial_j f_{\nu l}]_{\star_{3'}})
+(a_i\star_2\partial_j f_{\mu\nu})(a_k\star_2\partial_l f^{\mu\nu})
\\&
+\frac{1}{2}f^{\mu\nu}(2[a_i\partial_j a_k \partial_l f_{\mu\nu}]_{\star_{3'}}+2[\partial_l f_{\mu\nu}a_i \partial_j a_k]_{\star_{3'}}+2[a_i a_k \partial_j\partial_l f_{\mu\nu}]_{\star_{3'}}
-[a_i\partial_k a_j\partial_l f_{\mu\nu}]_{\star_{3'}}-[\partial_l f_{\mu\nu}a_i\partial_k a_j]_{\star_{3'}}),
\end{split}
\label{2.8}
\end{equation}
and
\begin{equation}
\begin{split}
S^{e^2}_{\rm (II)}=&-\frac{e^2}{4}\theta^{ij}\theta^{kl}\int\,\kappa^2(f_{\mu i}\star_2 f_{\nu j})(f^\mu_{\;\,\; k}\star_2 f^\nu_{\;\,\; l})-\kappa(f_{ij}\star_2 f_{\mu\nu})(f^\mu_{\;\,\; k}\star_2 f^\nu_{\;\,\; l})
\\&+2\kappa'_1 f^{\mu\nu}\left(2f_{\mu i}\star_2(f_{jk}\star_2 f_{l\nu})-[f_{\mu i}f_{jk}f_{l\nu}]_{\star_3}\right)
-4\kappa'_2 f^{\mu\nu}((a_i\star_2\partial_j f_{\mu k})\star_2 f_{\nu l}-[a_i\partial_j f_{\mu k}f_{\nu l}]_{\star_3})
\\&+(a_i\star_2\partial_j f_{\mu\nu})(a_k\star_2\partial_l f^{\mu\nu})
+f^{\mu\nu}(2 a_i\star_2(\partial_j a_k\star_2\partial_l f_{\mu\nu})+2(a_i\star_2\partial_j a_k)\star_2\partial_l f_{\mu\nu}-2[a_i\partial_j a_k\partial_l f_{\mu\nu}]_{\star_3}
\\&-a_i\star_2(\partial_k a_j\star_2\partial_l f_{\mu\nu})-(a_i\star_2\partial_k a_j)\star_2\partial_l f_{\mu\nu}+[a_i\partial_k a_j\partial_l f_{\mu\nu}]_{\star_3} +2a_i\star_2(a_k\star_2\partial_j\partial_l f_{\mu\nu})-[a_i a_k \partial_j\partial_l f_{\mu\nu}]_{\star_2}).
\end{split}
\label{2.9}
\end{equation}
One can reduce \eqref{2.8} and \eqref{2.9} to the leading/$\theta^2$ order and perform
integration-by-part to obtain
\begin{equation}
S^{\theta^2}_{\rm (I)}=S^{\theta^2}_{\rm (II)}=-\frac{e^2}{4}\int \kappa^2 f_{\mu i}f_{\nu j}f^\mu_{\;\,\; k}f^\nu_{\;\,\; l}-\kappa f^{\mu\nu}f_{\mu i}f_{\nu j}f_{kl}+2\kappa_1 f^{\mu\nu}f_{\mu i}f_{\nu k}f_{jl}-\frac{1}{4}f^{\mu\nu}f_{\mu\nu}f_{ik}f_{jl}+\frac{1}{8}f^{\mu\nu}f_{ij}f_{kl}f_{\mu\nu}.
\label{thetasquare}
\end{equation}
We observe two crucial facts from this formula: First, two more gauge invariant
structures $f^{\mu\nu}f_{\mu\nu}f_{ik}f_{jl}$ and $f^{\mu\nu}f_{ij}f_{kl}f_{\mu\nu}$ emerge
after the integration-by-part. Second, these five terms deplete all possible combination of
four $\rm U(1)$ field strength $f_{\mu\nu}$ contracted with two $\theta$ and two
metric tensors. One would naturally wonder what are the $\theta$-exact completions of these
two terms and if there are more structures emerging after a $\theta$-exact integration-by-part
procedure performed. This is, however, not an easy task, since the universal
integration-by-part/cyclicity identities on Moyal space only exist for the
two-products/integral over the product of three functions. Yet it is still possible to
convert both $\theta$-exact interactions \eqref{2.8} and \eqref{2.9} fully in terms of the
commutative field strength $f_{\mu\nu}$ by applying a series of integration-by-part
identities resulting from verification of  the Ward identity on the four-photon coupling
vertex. A detailed procedure for model I is given in Appendix A, while here we only list
the final result for both models I and II:
\begin{equation}
\begin{split}
S^{e^2}_{\rm (I)}=&-\frac{e^2}{4}\theta^{ij}\theta^{kl}\int\,\kappa^2(f_{\mu i}\star_2 f_{\nu j})(f^\mu_{\;\,\; k}\star_2 f^\nu_{\;\,\; l})-\kappa(f_{ij}\star_2 f_{\mu\nu})(f^\mu_{\;\,\; k}\star_2 f^\nu_{\;\,\; l})
\\&+2\kappa_1 f^{\mu\nu}[f_{\mu i}f_{\nu k}f_{jl}]_{\star_{3'}}+2\kappa_2 f^{\mu\nu}(a_i\star_2\partial_j(f_{\mu k}\star_2 f_{\nu l})-[f_{\mu k} a_i\partial_j f_{\nu l}]_{\star_{3'}}-[a_i f_{\mu k}\partial_j f_{\nu l}]_{\star_{3'}})
\\&-\frac{1}{4}f^{\mu\nu}\left[f_{\mu\nu}f_{ik}f_{jl}\right]_{\star_{3'}}+\frac{1}{8}\left(f^{\mu\nu}\star_2 f_{ij}\right)\left(f_{kl}\star_2 f_{\mu\nu}\right)+\frac{1}{2}\theta^{pq}f^{\mu\nu}\left[\partial_i f_{jk} f_{lp}\partial_q f_{\mu\nu}\right]_{\mathcal M_{\rm (I)}},
\end{split}
\label{2.10}
\end{equation}
and
\begin{equation}
\begin{split}
S^{e^2}_{\rm (II)}=&-\frac{e^2}{4}\theta^{ij}\theta^{kl}\int\,\kappa^2(f_{\mu i}\star_2 f_{\nu j})(f^\mu_{\;\,\; k}\star_2 f^\nu_{\;\,\; l})-\kappa(f_{ij}\star_2 f_{\mu\nu})(f^\mu_{\;\,\; k}\star_2 f^\nu_{\;\,\; l})
\\&+2\kappa'_1f^{\mu\nu}(2f_{\mu i}\star_2(f_{jk}\star_2 f_{l\nu})-[f_{\mu i}f_{jk}f_{l\nu}]_{\star_3})
-4\kappa'_2 f^{\mu\nu}((a_i\star_2\partial_j f_{\mu k})\star_2 f_{\nu l}-[a_i\partial_j f_{\mu k}f_{\nu l}]_{\star_3})
\\&-\frac{1}{4}f^{\mu\nu}(3f_{ik}\star_2\left(f_{jl}\star_2 f_{\mu\nu}\right)-2\left[f_{ik}f_{jl}f_{\mu\nu}\right]_{\star_3})
+\frac{1}{8}f^{\mu\nu}(2f_{ij}\star_2\left(f_{kl}\star_2 f_{\mu\nu}\right)-\left[f_{ij}f_{kl}f_{\mu\nu}\right]_{\star_3})
\\&-\frac{1}{4}\theta^{pq}\theta^{rs}f^{\mu\nu}\left[\partial_k f_{ri}\partial_j f_{lp}\partial_q\partial_s f_{\mu\nu}+\partial_i\partial_r f_{jk}\partial_s(f_{lp}\partial_q f_{\mu\nu})\right]_{\mathcal M_{\rm (II)}}.
\end{split}
\label{2.11}
\end{equation}
The products $\mathcal M_{\rm (I,II)}$ are defined via the
momentum structures $f_{\rm (I,II)}$ in Appendix A.

As stated above, the five terms of order $\theta^2$ deplete all possible indices
arrangements. Also $\theta^{ij}\theta^{kl}f_{ik}f_{jl}f_{\mu\nu}$ and
$\theta^{ij}\theta^{kl}f_{ij}f_{kl}f_{\mu\nu}$ terms can be easily generated via the $\theta$-iterative
procedure ~\cite{Bichl:2001cq}. Therefore it is reasonable to introduce two additional freedom
parameters $(\kappa_3, \kappa_4)$ and $(\kappa'_3, \kappa'_4)$, in (I) and (II) respectively, as the $\theta$-exact completion of these two freedoms. In this way we  produce the final forms for the $a_{\mu}^4$-order actions (I,II):
\begin{equation}
\begin{split}
S^{e^2}_{\rm (I)_{\kappa,\kappa_1,\kappa_2,\kappa_3,\kappa_4}}=&-\frac{e^2}{4}\theta^{ij}\theta^{kl}\int\,\kappa^2(f_{\mu i}\star_2 f_{\nu j})(f^\mu_{\;\,\; k}\star_2 f^\nu_{\;\,\; l})-\kappa(f_{ij}\star_2 f_{\mu\nu})(f^\mu_{\;\,\; k}\star_2 f^\nu_{\;\,\; l})
\\&+2\kappa_1f^{\mu\nu}[f_{\mu i}f_{\nu k}f_{jl}]_{\star_{3'}}+2\kappa_2 f^{\mu\nu}
(a_i\star_2\partial_j(f_{\mu k}\star_2 f_{\nu l})-[f_{\mu k} a_i\partial_j f_{\nu l}]_{\star_{3'}}-[a_i f_{\mu k}\partial_j f_{\nu l}]_{\star_{3'}})
\\&
-\frac{\kappa_3}{4}f^{\mu\nu}\left[f_{\mu\nu}f_{ik}f_{jl}\right]_{\star_{3'}}+\frac{\kappa_4}{8}\left(f^{\mu\nu}\star_2 f_{ij}\right)\left(f_{kl}\star_2 f_{\mu\nu}\right)+\frac{1}{2}\theta^{pq} f^{\mu\nu} \left[\partial_i f_{jk} f_{lp}\partial_q f_{\mu\nu}\right]_{\mathcal M_{\rm (I)}},
\end{split}
\label{2.15}
\end{equation}
and
\begin{equation}
\begin{split}
S^{e^2}_{\rm (II)_{\kappa,\kappa'_1,\kappa'_2,\kappa'_3,\kappa'_4}}=&-\frac{e^2}{4}\theta^{ij}\theta^{kl}\int\,\kappa^2(f_{\mu i}\star_2 f_{\nu j})(f^\mu_{\;\,\; k}\star_2 f^\nu_{\;\,\; l})-\kappa(f_{ij}\star_2 f_{\mu\nu})(f^\mu_{\;\,\; k}\star_2 f^\nu_{\;\,\; l})
\\&+2\kappa'_1f^{\mu\nu}(2f_{\mu i}\star_2(f_{jk}\star_2 f_{l\nu})-[f_{\mu i}f_{jk}f_{l\nu}]_{\star_3})
-4\kappa'_2f^{\mu\nu}((a_i\star_2\partial_j f_{\mu k})\star_2 f_{\nu l}-[a_i\partial_j f_{\mu k}f_{\nu l}]_{\star_3})
\\&-\frac{\kappa'_3}{4}f^{\mu\nu}\left(3f_{ik}\star_2(f_{jl}\star_2 f_{\mu\nu}\right)-2\left[f_{ik}f_{jl}f_{\mu\nu}\right]_{\star_3})
+\frac{\kappa'_4}{8}f^{\mu\nu}\left(2f_{ij}\star_2(f_{kl}\star_2 f_{\mu\nu}\right)-\left[f_{ij}f_{kl}f_{\mu\nu}\right]_{\star_3})
\\&-\frac{1}{4}\theta^{pq}\theta^{rs}f^{\mu\nu}\left[\partial_k f_{ri}\partial_j f_{lp}\partial_q\partial_s f_{\mu\nu}+\partial_i\partial_r f_{jk}\partial_s(f_{lp}\partial_q f_{\mu\nu})\right]_{\mathcal M_{\rm (II)}}.
\end{split}
\label{2.16}
\end{equation}
Note that $\kappa$-terms are identical for both above actions, as they should be in the $\kappa$-settings.

\subsection{Feynman rule for the four-photon interaction}

\begin{figure}
\begin{center}
\includegraphics[width=6cm,height=4cm]{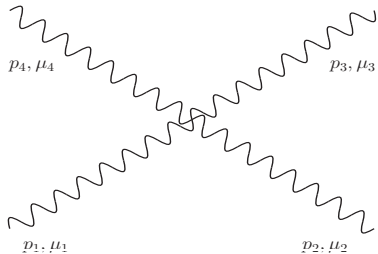}
\end{center}
\caption{Four-photon field vertex $\Gamma^{\mu_1\mu_2\mu_3\mu_4}(p_1,p_2,p_3,p_4)$ with all incoming momenta.}
\label{fig:vertex}
\end{figure}

Since the triple-photon Feynman rules from action (\ref{3photon})
is given previously in \cite{Trampetic:2014dea} it is not necessary
to be repeated here.
From the actions \eqref{2.15} and \eqref{2.16} we read out the corresponding
four-photon interactions in the momentum space, with all four
momenta $p_i$ in Fig.\ref{fig:vertex} being the incoming ones
\begin{equation}
\begin{split}
\Gamma^{\mu_1\mu_2\mu_3\mu_4}_{\rm (I)}\left(p_1,p_2,p_3,p_4\right)=&-i\frac{e^2}{4}[(\kappa^2\Gamma_{\rm A}^{\mu_1\mu_2\mu_3\mu_4}\left(p_1,p_2,p_3,p_4\right)
+\kappa\Gamma_{\rm B}^{\mu_1\mu_2\mu_3\mu_4}\left(p_1,p_2,p_3,p_4\right)
\\&+\kappa_1\Gamma_1^{\mu_1\mu_2\mu_3\mu_4}\left(p_1,p_2,p_3,p_4\right)+\kappa_2\Gamma_2^{\mu_1\mu_2\mu_3\mu_4}\left(p_1,p_2,p_3,p_4\right)
+\kappa_3\Gamma_3^{\mu_1\mu_2\mu_3\mu_4}\left(p_1,p_2,p_3,p_4\right)
\\&+\kappa_4\Gamma_4^{\mu_1\mu_2\mu_3\mu_4}\left(p_1,p_2,p_3,p_4\right)+\Gamma_5^{\mu_1\mu_2\mu_3\mu_4}\left(p_1,p_2,p_3,p_4\right))
\\&+{\rm all\; S_4\; permutations\; over}\; \{p_i\}{\rm \; and}\; \{\mu_i\}{\rm \; simutaneously}]\delta\left(p_1+p_2+p_3+p_4\right).
\end{split}
\label{3.1}
\end{equation}
and
\begin{equation}
\begin{split}
\Gamma^{\mu_1\mu_2\mu_3\mu_4}_{\rm (II)}\left(p_1,p_2,p_3,p_4\right)=&-i\frac{e^2}{4}[(\kappa^2\Gamma_{\rm A}^{\mu_1\mu_2\mu_3\mu_4}\left(p_1,p_2,p_3,p_4\right)
+\kappa\Gamma_{\rm B}^{\mu_1\mu_2\mu_3\mu_4}\left(p_1,p_2,p_3,p_4\right)
\\&+\kappa'_1{\Gamma'}_1^{\mu_1\mu_2\mu_3\mu_4}\left(p_1,p_2,p_3,p_4\right)+\kappa'_2{\Gamma'}_2^{\mu_1\mu_2\mu_3\mu_4}\left(p_1,p_2,p_3,p_4\right)
+\kappa'_3{\Gamma'}_3^{\mu_1\mu_2\mu_3\mu_4}\left(p_1,p_2,p_3,p_4\right)
\\&+\kappa'_4{\Gamma'}_4^{\mu_1\mu_2\mu_3\mu_4}\left(p_1,p_2,p_3,p_4\right)+{\Gamma'}_5^{\mu_1\mu_2\mu_3\mu_4}\left(p_1,p_2,p_3,p_4\right))
\\&+{\rm all\; S_4\; permutations\; over}\; \{p_i\}{\rm \; and}\; \{\mu_i\}{\rm \; simutaneously}]\delta\left(p_1+p_2+p_3+p_4\right).
\end{split}
\label{3.2}
\end{equation}
with $\Gamma_{\rm A}$, $\Gamma_{\rm B}$, $\Gamma_i$ and $\Gamma'_i$ 's being given in the Appendix B.

\section{Four-photon interaction at one loop}

\begin{figure}
\begin{center}
\includegraphics[width=7cm,height=4cm]{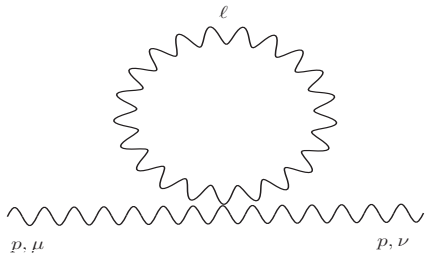}
\end{center}
\caption{Four-photon-tadpole contribution to the photon two-point function $T^{\mu\nu}(p)$.}
\label{fig:photontadpole}
\end{figure}

Following the successful derivation of four-photon self-coupling vertices we move on to
study the simplest perturbative quantum loop correction induced by this coupling: the one-loop
four-photon-tadpole diagram contribution to the photon polarization tensor. We first read
off from Fig.\ref{fig:photontadpole} the tadpole integral. After some arithmetics we find that various nonlocal factors listed in the Appendix A simplify into two cases: either one or $f_{\star_2}^2(p,\ell)$, i.e.
\begin{equation}
\begin{split}
T^{\mu\nu}_{\rm (I,II)}(p)=&\frac{1}{2}\mu^{d-D}\int\,\frac{d^D \ell}{(2\pi)^D}
\frac{-ig_{\rho\sigma}}{\ell^2}\Gamma^{\mu\nu\rho\sigma}_{\rm (I,II)}(p,-p,\ell,-\ell)\\
=&e^2\tau^{\mu\nu}_{\rm (I,II)}\;\mu^{d-D} \int\,\frac{d^D \ell}{(2\pi)^D}\frac{\ell^2}{\ell^2}+e^2\mathcal T_{\rm (I,II)}^{\mu\nu\rho\sigma}\mu^{d-D} \int\,\frac{d^D \ell}{(2\pi)^D}\frac{\ell_\rho \ell_\sigma}{\ell^2}f_{\star_2}^2(p,\ell).
\end{split}
\label{3.3}
\end{equation}
Since the first integral in the above Eq. (\ref{3.3}) vanishes according to the dimensional regularization prescription~\cite{Leibbrandt:1975dj}, the only remaining integral is the second one. The tensors $\tau^{\mu\nu}_{\rm (I,II)}$ and ${\cal T}_{\rm (I,II)}^{\mu\nu\rho\sigma}$ are given in the Appendix C. 
\subsection{Tadpole integral}

Now we compute our $D$-dimensional tadpole integral
 \begin{equation}
I^{\mu\nu}=\int\,\frac{d^D \ell}{(2\pi)^D}\frac{\ell^\mu \ell^\nu}{\ell^2}f_{\star_2}^2(p,\ell).
\label{3.4}
\end{equation}
We have encounter similar computation in our prior works~\cite{Horvat:2011qn,Horvat:2013rga}. Here to ensure the consistency of our methodology we choose to evaluate the integral in two different ways, both are extensions of sound methods for commutative field theories. As shown below these two methods agrees with each other at the $D\to 4$ limit, as expected.

(1) The first is the conventional method which multiplies the tadpole integral with an identity $1=(\ell+p)^2/(\ell+p)^2$ to turn it into a bubble integral. We then parameterizing the integral as in our prior works \cite{Horvat:2013rga}, which ultimately yields the following expression in $D$ dimensions
\begin{equation}
\begin{split}
&I^{\mu\nu}=g^{\mu\nu}\frac{1}{D-1}\left(4(\theta p)^{-\frac{D}{2}}\mathcal K\left[\frac{D}{2}-1;0,0\right]-\frac{4p^2}{(\theta p)^2}\left((1-D)\mathcal K\left[\frac{D}{2}-2;0,1\right]
\right.\right.\\&\left.\left.+2D\mathcal K\left[\frac{D}{2}-2;1,1\right]\right)-p^2\left((1-D)\mathcal W\left[\frac{D}{2}-1;0,1\right]-2D\mathcal W\left[\frac{D}{2}-1;1,1\right]\right)\right)
\\&+p^\mu p^\nu\left(\frac{4}{(\theta p)^2}\left((D-2)\mathcal K\left[\frac{D}{2}-2;1,0\right]+2(1-D)\mathcal K\left[\frac{D}{2}-2;1,1\right]\right)\right.\\&\left.-\left((1-D)\mathcal W\left[\frac{D}{2}-1;1,0\right]+2D\mathcal W\left[\frac{D}{2}-1;1,1\right]\right)\right)
\\&+(\theta p)^\mu(\theta p)^\nu\frac{1}{D-1}\left(-4D(\theta p)^{-1-\frac{D}{2}}\mathcal K\left[\frac{D}{2}-1;0,0\right]+\frac{4p^2}{(\theta p)^4}\left((1-D)\mathcal K\left[\frac{D}{2}-2;0,1\right]
\right.\right.\\&\left.\left.+(2D-1)\mathcal K\left[\frac{D}{2}-2;1,1\right]\right)+\frac{p^2}{(\theta p)^2}\left((1-D)\mathcal W\left[\frac{D}{2}-1;0,1\right]+2D\mathcal W\left[\frac{D}{2}-1;1,1\right]\right)\right),
\end{split}
\label{3.5}
\end{equation}
where the special function integrals are defined as follows
\begin{gather}
\mathcal{K}[\nu;a,b]=2^{\nu}(\theta p)^{-\nu}\int\limits_0^1\,dx\,x^a(1-x)^b  X^{\nu}  K_\nu[X],
\label{3.6}\\
\mathcal{W}[\nu;a,b]=\int\limits_0^1\,dx\,x^a(1-x)^b W_\nu[X].
\label{3.7}
\end{gather}
Here the $K_\nu[X]$ is the modified Bessel function of second kind, while
\begin{eqnarray}
W_\nu[X]=
(\theta p)^{-2\nu}\Bigg[X^{2\nu}\Gamma\left[-\nu\right] {}_1F_2\left(\frac{1}{2};\frac{3}{2},\nu+1;\frac{X^2}{4}\right)
-\frac{2^{2\nu}}{1-2\nu}\Gamma\left[\nu\right]{}_1F_2\left(\frac{1-2\nu}{2};1-\nu,\frac{3-2\nu}{2};\frac{X^2}{4}\right)\Bigg],
\label{3.8}
\end{eqnarray}
and
\begin{equation}
X=(x(1-x)p^2(\theta p)^2)^{\frac{1}{2}}.
\label{3.9}
\end{equation}
In the limit $D\to 4-\epsilon$ after a lengthy manipulations the tadpole integral $I^{\mu\nu}$ reduces to the following two terms
\begin{equation}
I^{\mu\nu}=\frac{1}{6\pi^2}\frac{1}{(\theta p)^4}\left(g^{\mu\nu}-4\frac{(\theta p)^\mu(\theta p)^\nu}{(\theta p)^2}\right).
\label{3.10}
\end{equation}
The integrals $\mathcal K\left[\frac{D}{2}-2;0,1\right]$ and $\mathcal K\left[\frac{D}{2}-2;1,0\right]$ become UV divergent when $D\to 2-\epsilon$. This
hard $1/\epsilon$ divergence occurs at the the same place as the IR divergence when $D\to 4-\epsilon$; i.e., the UV and IR divergence get "mixed" in the same term now.

(2) As a test a second computation using the $n$-nested zero regulator~\cite{Leibbrandt:1975dj} is also performed. In this approach we first parameterize the integral as follows
\begin{equation}
\begin{split}
I^{\mu\nu}=
\int\,\frac{d^D \ell}{(2\pi)^D}\frac{\ell^\mu \ell^\nu}{\ell^2}f_{\star_2}^2(p,\ell)
=-2\int\limits_0^\infty d\lambda\; \lambda^2
\,\int\limits_0^{1/\lambda} dy \left(y-\frac{1}{\lambda}\right)\,
\\
\cdot\int\,\frac{d^D \ell}{(2\pi)^D}\left(\frac{\ell^2}{D}g^{\mu\nu}-\frac{y^2}{4}(\theta p)^\mu(\theta p)^\nu\right)\exp\left[-\lambda \ell^2-\frac{\lambda y^2}{4}\left(\theta p\right)^2\right].
\end{split}
\label{3.11}
\end{equation}
Next we introduce the loop momenta integral
\begin{equation}
\int\,\frac{d^D \ell}{(2\pi)^D}\exp\left[-\lambda \ell^2\right]=\left(4\pi\lambda\right)^{-\frac{D}{2}}\exp\left[-\lambda f\left(\frac{D}{2}\right)\right],
\label{3.12}
\end{equation}
where $f\left(\frac{D}{2}\right)$ is the $n$-nested zero regulator~\cite{Leibbrandt:1975dj} which satisfies the following properties
\begin{itemize}
\item $f\left(\frac{D}{2}\right)$ is a nonzero analytic function
\item $f\left(\frac{D}{2}\right)=0$ when $D\in Z^+$
\item $f^{(\ell)}\left(\frac{D}{2}\right)=0$ when $D\in Z^+$ and $\ell\le \ell_0\in N$
\item $\forall {\rm Re}D\not\in Z^+,\; \exists {\rm Im} D,\; {\rm Re}\left[f\left(\frac{D}{2}\right)\right]>0$.
\end{itemize}
The inclusion of the $f\left(\frac{D}{2}\right)$ regulator leads to a planar-modified Bessel-hypergeometric function combinations similar to the first approach with but a variable $\left(f\left(\frac{D}{2}\right)(\theta p)^2\right)^{\frac{1}{2}}$, instead of $\left(x(1-x)p^2(\theta p)^2\right)^{\frac{1}{2}}$, and no $x$ integration. In the
end, we have found
\begin{equation}
I^{\mu\nu}=I\cdot\left(-\frac{g^{\mu\nu}}{D}+\frac{(\theta p)^\mu(\theta p)^\nu}{(\theta p)^2}\right),
\label{3.13}
\end{equation}
with
\begin{equation}
\begin{split}
I=(4\pi)^{-\frac{D}{2}}&\left(\frac{(\theta p)^2}{4}\right)^{-1}\left\{\left(f\left(\frac{D}{2}\right)\right)^{\frac{D}{2}-1}\Gamma\left(1-\frac{D}{2}\right)-W_{\frac{D}{2}}\left[\left(f\left(\frac{D}{2}\right)(\theta p)^2\right)^{\frac{1}{2}}\right]\right.
\\&\left.-2\left(f\left(\frac{D}{2}\right)\right)^{\frac{D}{4}-\frac{1}{2}}\left(\frac{(\theta p)^2}{4}\right)^{-\frac{1}{2}-\frac{D}{4}}K_{\frac{D}{2}-1}\left[\left(f\left(\frac{D}{2}\right)(\theta p)^2\right)^{\frac{1}{2}}\right]\right\}.
\end{split}
\label{3.14}
\end{equation}
The above scalar integral $I$ reduces to a single IR divergent term
$-\frac{2}{3\pi^2}(\theta p)^{-4}$ when $D\to 4-\epsilon$, with all others
being suppressed by the $f\left(\frac{D}{2}\right)$ regulator. This matches
the result (\ref{3.10}) obtained from the other method. 

\subsection{Four-photon-tadpole contributions in the limit $D\to 4-\epsilon$}

By combining the partial tensor reduction from above and the master integral in $D=4$, we obtain the following results
in the $\kappa$-settings,
\begin{equation}
\begin{split}
T_{\rm (I,II)}^{\mu\nu}(p)&=\frac{e^2}{(4\pi)^2}\{[g^{\mu\nu}p^2-p^\mu p^\nu]T_{1_{\rm (I,II)}}(p)
+(\theta p)^\mu (\theta p)^\nu T_{2_{\rm (I,II)}}(p)
+[g^{\mu\nu}(\theta p)^2-(\theta\theta)^{\mu\nu}p^2
+ p^{\{\mu}(\theta\theta p)^{\nu\}}]T_{3_{\rm (I,II)}}(p)
\\&+[(\theta\theta)^{\mu\nu}(\theta p)^2+(\theta\theta p)^\mu(\theta\theta p)^\nu]T_{4_{\rm (I,II)}}(p)
+ (\theta p)^{\{\mu} (\theta\theta\theta p)^{\nu\}} T_{5_{\rm (I,II)}}(p)\},
\end{split}
\label{4.1T}
\end{equation}
with model I coefficients
\begin{equation}
\begin{split}
T_{1_{\rm (I)}}(p)&=\frac{4}{3}\left(\frac{tr\theta\theta}{(\theta p)^4}+4\frac{(\theta\theta p)^2}{(\theta p)^6}\right)\big(\kappa_4-1\big),
\\
T_{2_{\rm (I)}}(p)&=\frac{16}{3}\frac{1}{(\theta p)^4} \,\big(2\kappa^2-4\kappa+6\kappa_1+2\kappa_2-2\kappa_3+\kappa_4-1\big),
\\
T_{3_{\rm (I)}}(p)&=\frac{16}{3}\frac{1}{(\theta p)^4}\big(2\kappa^2-2\kappa+\kappa_1+\kappa_2\big),
\\
T_{4_{\rm (I)}}(p)&=\frac{32}{3}\frac{p^2}{(\theta p)^6}\big(\kappa^2-2\kappa+\kappa_1+\kappa_2\big),
\\
T_{5_{\rm (I)}}(p)&=\frac{16}{3}\frac{p^2}{(\theta p)^6}\big(2\kappa-\kappa_1-\kappa_2\big),
\end{split}
\label{Tik}
\end{equation}
and model II
\begin{equation}
\begin{split}
T_{1_{\rm (II)}}(p)&=\frac{4}{3}\left(\frac{tr\theta\theta}{(\theta p)^4}+4\frac{(\theta\theta p)^2}{(\theta p)^6}\right)\big(2\kappa'_2+\kappa'_4-3\big),
\\
T_{2_{\rm (II)}}(p)&=\frac{32}{3}\frac{1}{(\theta p)^4} \,\big(\kappa^2-2\kappa+2\kappa'_1+2\kappa'_2-\kappa'_3\big),
\\
T_{3_{\rm (II)}}(p)&=\frac{16}{3}\frac{1}{(\theta p)^4}\big(2\kappa^2-2\kappa+2\kappa'_2\big),
\\
T_{4_{\rm (II)}}(p)&=\frac{32}{3}\frac{p^2}{(\theta p)^6}\big(\kappa^2-2\kappa+2\kappa'_2\big),
\\
T_{5_{\rm (II)}}(p)&=\frac{32}{3}\frac{p^2}{(\theta p)^6}\big(\kappa-\kappa'_2\big).
\end{split}
\label{Tik'}
\end{equation}
The tensor structure remains exactly the same as for the photon bubble diagram (Fig. \ref{fig:photonbubble}), as one would expect. However, we notice immediately the absence of UV and logarithmic divergent terms contrary to the photon bubble diagram results \cite{Horvat:2013rga}. In addition, the tadpole diagram produces no finite terms either. The tadpole contribution for model I can be made equal to that of model II when setting
\begin{eqnarray}
\kappa_1+\kappa_2&=&2\kappa'_2,
\nonumber\\
4\kappa_1-2\kappa_3&=&4\kappa'_1-4\kappa'_3-\kappa'_4+3,
\nonumber\\
\kappa_4&=&2\kappa'_3+\kappa'_4-2,
\label{3.21}
\end{eqnarray}
and in particular $T^{\mu\nu}_{\rm (I)}(p)=T^{\mu\nu}_{\rm (II)}(p)$, for $\kappa_i=\kappa'_i=1, \forall i$. We also notice that $T_3$, $T_4$, and $T_5$ can only be set to zero simultaneously if $\kappa=0$. Since the value $\kappa$ directly affects the divergences in the bubble diagram~\cite{Horvat:2013rga}, we therefore conclude that the tadpole divergences
should  be analyzed only after being summed with the bubble diagram contribution.

\section{Summing over bubble and tadpole diagrams for $D\to 4-\epsilon$}

\begin{figure}
\begin{center}
\includegraphics[width=8cm,height=4.5cm]{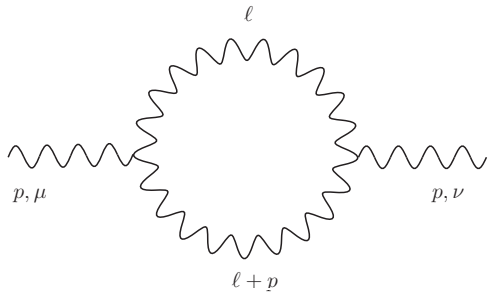}
\end{center}
\caption{Three-photon-bubble contribution to the photon two-point function $B^{\mu\nu}(p)$.}
\label{fig:photonbubble}
\end{figure}

As stated above to complete our analysis for the photon two-point function we have to sum up the tadpole (Fig. \ref{fig:photontadpole}) (\ref{3.3}) and the bubble (Fig. \ref{fig:photonbubble}) contributions, which has the same structure as (\ref{4.1T}), but with the coefficients $T_i(p)$ replaced with $B_i(p)$ given below.
To do that we express the bubble contribution ${\cal B}^{\mu\nu}(p)$, in terms of the $\kappa$-settings for the gauge field strength at the $e^2$ order (\ref{2.1}), by converting each $B_i$-coefficient from the $\kappa_g$-setting~\cite{Horvat:2013rga,Trampetic:2014dea} to the $\kappa$-setting. We first substitute $\kappa_g=\kappa^{-1}$, then extract out from all coefficients a factor $\kappa^{-2}$ since it is absorbed as an overall re-scaling factor of the corresponding field redefinition~\cite{Trampetic:2014dea}.  Remaining divergent parts of the $B_i$-coefficients are then given next in (\ref{4.2}) in terms of the $\kappa$-settings, so that they match their tadpole counterparts and could be summed up with.
\begin{gather}
\begin{split}
B_1(p)&\sim+\bigg(\frac{1}{3}\big(1-3\kappa\big)^2 +\frac{1}{3}\big(1+2\kappa\big)^2\; \frac{p^2(\tr\theta\theta)}{(\theta p)^2}
+\frac{2}{3}\big(1+4\kappa+\kappa^2\big)\; \frac{p^2(\theta\theta p)^2}{(\theta p)^4}\bigg)
\left[\frac{2}{\epsilon} + \ln(\mu^2(\theta p)^2)\right]
\\&-\frac{8}{3}\big(1-\kappa\big)^2\;\frac{1}{(\theta p)^6}\bigg((\tr\theta\theta)(\theta p)^2+4(\theta\theta p)^2\bigg)\,,
\\
B_2(p)&\sim+\bigg(\frac{4}{3}\big(1-\kappa\big)^2\; \frac{p^4(\theta\theta p)^2}{(\theta p)^6}+\frac{1}{3}\big(1-2\kappa-5\kappa^2\big)\frac{p^4(\tr\theta\theta)}{(\theta p)^4}+\frac{1}{3}\big(25
-86\kappa+73\kappa^2\big)\frac{p^2}{(\theta p)^2}\bigg)\left[\frac{2}{\epsilon} + \ln(\mu^2(\theta p)^2)\right]
\\&-\frac{8}{3}\big(1-3\kappa\big)\big(3-\kappa\big)\frac{1}{(\theta p)^4}
+\frac{16}{3}(1-\kappa\big)^2\frac{p^2}{(\theta p)^8}\bigg((\tr\theta\theta)(\theta p)^2+6(\theta\theta p)^2\bigg),
\\
B_3(p)&\sim-\frac{1}{6}\big(1-2\kappa-11\kappa^2\big)\frac{p^2}{(\theta p)^2}
\left[\frac{2}{\epsilon} + \ln(\mu^2(\theta p)^2)\right]
-\frac{4}{3(\theta p)^4}\big(1-10\kappa+17\kappa^2\big),
\\
B_4(p)&\sim-\big(1+\kappa\big)^2\frac{p^4}{(\theta p)^4}\left[\frac{2}{\epsilon} + \ln(\mu^2(\theta p)^2)\right]-\frac{16p^2}{3(\theta p)^6}\big(1-6\kappa+7\kappa^2\big),
\\
B_5(p)&\sim+\frac{2}{3}\big(1+\kappa+4\kappa^2\big)\frac{p^4}{(\theta p)^4}
\left[\frac{2}{\epsilon} + \ln(\mu^2(\theta p)^2)\right]+\frac{32p^2}{3(\theta p)^6}\big(1-\kappa\big)\big(1-2\kappa\big).
\end{split}
\label{4.2}
\end{gather}
All $B_i(p)$ coefficients are computed for arbitrary $\kappa$ and $\sim$ means that we have neglected all finite terms in the above equations. We observe the presence of the UV divergences as well as quadratic UV/IR mixing in all $B_i$'s. The logarithmic IR divergences from both planar and nonplanar sources in the bubble diagram appear to have identical coefficient and combine into a single $\ln\mu^2(\theta p)^2$ term. Finally, no single $\kappa$ value is capable of removing all novel divergences. 

\subsection{$\theta$ independent elimination of the bubble plus tadpole IR divergences}

Since  tadpole integrals give only a quadratic IR divergence, we perform a sum
over the tadpole and quadratically IR divergent parts of the bubble diagram~\cite{Horvat:2013rga} for an arbitrary choice of the antisymmetric tensor $\theta^{\mu\nu}$  and deformation freedom parameters. Working out the arithmetics we get for both models, I and II in the IR regime the sum $\Pi_{\rm(I,II)}^{\mu\nu}(p)_{\rm IR}={\cal B}^{\mu\nu}(p)_{\rm IR} + T^{\mu\nu}_{\rm (I,II)}(p)_{\rm IR}$ of the bubble and the tadpole photon contribution to the photon polarization tensor. It has again the structure as in (\ref{4.1T}), but with coefficients $T_i(p)$ getting replaced with the following sums:
\begin{equation}
\Pi_{i_{\rm (I,II)}}(p)_{\rm IR}=B_i(p)_{\rm IR}+T_{i_{\rm (I,II)}}(p)_{\rm IR}, \;\forall i=1,...,5.
\label{4.5IR}
\end{equation}
Since tadpole produces no UV, log and also has no finite contributions, we have
\begin{equation}
\Pi_i(p)_{\rm UV}=B_i(p)_{\rm UV}, \;\forall i=1,...,5.
\label{4.5UV}
\end{equation}
According to (\ref{4.5IR}), in the rest of this article label $\Pi^{\mu\nu}$ always represents the sum of bubble and tadpole contributions to the photon polarization tensor. A summation over the leading IR terms in the bubble and tadpole diagrams provides IR results (for overall UV/IR mixings) in model I:
\begin{gather}
\begin{split}
\Pi_{1_{\rm (I)}}(p)_{\rm IR}&\sim-\frac{4}{3}\frac{1}{(\theta p)^4}\Big(tr\theta\theta+4\frac{(\theta\theta p)^2}{(\theta p)^2}\Big)\Big(2\big(\kappa-1\big)^2-\kappa_4+1\Big),
\\
\Pi_{2_{\rm (I)}}(p)_{\rm IR}&\sim+\frac{8}{3}\frac{1}{(\theta p)^4}\Big(\kappa^2+2\kappa+12\kappa_1+4\kappa_2-4\kappa_3+2\kappa_4-5\Big)
+\frac{16}{3}\frac{p^2}{(\theta p)^4}\Big(\frac{tr\theta\theta}{(\theta p)^2}+6\frac{(\theta\theta p)^2}{(\theta p)^4}\Big)\Big(\kappa-1\Big)^2,
\\
\Pi_{3_{\rm (I)}}(p)_{\rm IR}&\sim-\frac{4}{3}\frac{1}{(\theta p)^4}\Big(9\kappa^2-2\kappa-4\kappa_1-4\kappa_2+1\Big),
\\
\Pi_{4_{\rm (I)}}(p)_{\rm IR}&\sim-\frac{16}{3}\frac{p^2}{(\theta p)^6}\Big(5\kappa^2-2\kappa-2\kappa_1-2\kappa_2+1\Big),
\\
\Pi_{5_{\rm (I)}}(p)_{\rm IR}&\sim+\frac{16}{3}\frac{p^2}{(\theta p)^6}\Big(4\kappa^2-4\kappa-\kappa_1-\kappa_2+2\Big),
\end{split}
\label{4.6}
\end{gather}
and in model II:
\begin{gather}
\begin{split}
\Pi_{1_{\rm (II)}}(p)_{\rm IR}&\sim-\frac{4}{3}\frac{1}{(\theta p)^4}\Big(tr\theta\theta+4\frac{(\theta\theta p)^2}{(\theta p)^2}\Big)\Big(2\big(\kappa-1\big)^2-2\kappa'_3-\kappa'_4+3\Big),
\\
\Pi_{2_{\rm (II)}}(p)_{\rm IR}&\sim+\frac{8}{3}\frac{1}{(\theta p)^4}\Big(\kappa^2+2\kappa+8\kappa'_1+8\kappa'_2-4\kappa'_3-3\Big)
+\frac{16}{3}\frac{p^2}{(\theta p)^4}\Big(\frac{tr\theta\theta}{(\theta p)^2}+6\frac{(\theta\theta p)^2}{(\theta p)^4}\Big)\Big(\kappa-1\Big)^2,
\\
\Pi_{3_{\rm (II)}}(p)_{\rm IR}&\sim-\frac{4}{3}\frac{1}{(\theta p)^4}\Big(9\kappa^2-2\kappa-8\kappa'_2+1\Big),
\\
\Pi_{4_{\rm (II)}}(p)_{\rm IR}&\sim-\frac{16}{3}\frac{p^2}{(\theta p)^6}\Big(5\kappa^2-2\kappa-4\kappa'_2+1\Big),
\\
\Pi_{5_{\rm (II)}}(p)_{\rm IR}&\sim+\frac{16}{3}\frac{p^2}{(\theta p)^6}\Big(4\kappa^2-4\kappa-2\kappa'_2+2\Big).
\end{split}
\label{4.11}
\end{gather}
Here we see that IR divergence in the coefficients $\Pi_{3,4,5}$ depends only
on $\kappa_1+\kappa_2$ in model I or $\kappa'_2$ in model II.  A bit more calculation shows that $\kappa=\kappa_1=\kappa_2=\kappa'_2=1$ can annihilate $\Pi_{3,4,5_{\rm (I,II)}}(p)_{\rm IR}$, respectively. Furthermore, $\kappa=1$ also removes the pathological second term in $\Pi_{2_{\rm (I,II)}}(p)_{\rm IR}$. The rest of divergences in \eqref{4.6} and  \eqref{4.11} can be readily removed by an appropriate choice of the rest of freedom parameters, proving
that the elimination of all IR divergences is independent on the choices of $\theta$-matrix elements.

\subsection{UV divergences}

Up to now we see that the quadratic IR divergence can be canceled by selecting the first order gauge freedom parameter $\kappa=1$ and $\forall \theta$. On the other hand, we also know that there are various UV ($1/\epsilon$) divergences in the bubble diagram which are also precisely connected to the logarithmic divergences. Characterizing this part of the behavior requires some simplifications by special $\theta^{\mu\nu}$ choices which satisfy the condition that $(\theta\theta)^{\mu\nu}$ becomes minus identity within a suitable subspace of dimension $n\le 4$. Here we consider two important examples.

\subsubsection{The subdimension  $n=2$ and $\theta^{\mu\nu}_1$ matrix}

First choice is to set $n=2$, as used already in \cite{Horvat:2011bs}.
This choice has the potential of preserving unitarity when the NC directions are spatial,
and it is manifested in the form of $\theta^{\mu\nu}_1$ matrix\footnote{Actually for $d=4$ any spacelike $\theta^{ij}$ can be simplified to this form by a rotation that sends the pseudovector $v_i=\epsilon_{ijk}\theta^{jk}$ to the third axis. Here $\epsilon_{ijk}$ is the totally antisymmetric tensor of the three spatial dimensions.}:
\begin{equation}
\theta^{\mu\nu}\equiv
\theta^{\mu\nu}_{1}=-\theta^2
\begin{pmatrix}
0&0&0&0\\
0&0&-1&0\\
0&1&0&0\\
0&0&0&0
\end{pmatrix}.
\label{nondegen1}
\end{equation}
So, for $d=4$-dimensional NC field theory and the subspace of dimension $n=2$, and by noticing that $(\theta\theta)^{\mu\nu}$ is now a projector into a two-dimensional subspace, we find
\begin{eqnarray}
-2\theta^2(\theta p)^\mu(\theta p)^\nu&=&2\Big[(\theta\theta)^{\mu\nu}(\theta p)^2+(\theta\theta p)^\mu(\theta\theta p)^\nu\Big]=(\theta p)^{\{\mu} (\theta\theta\theta p)^{\nu\}},
\label{theta1}\\
(\theta\theta\theta p)^\mu&=&-\theta^2 ({\theta p})^\mu, \:(\theta p)^2\tr\theta\theta=-2(\theta\theta p)^2=-2(\theta p)^2 \theta^2, \forall p;\; \theta^2=1/\Lambda_{\rm NC}^4,
\nonumber
\end{eqnarray}
with $\Lambda_{\rm NC}$ being the scale of noncommutativity.
In order to perform computations in this case, we need to use all above relations
in general decomposition of the photon polarization tensor, which is the same for both, the bubble and the tadpole contribution.
A general decomposition, like (\ref{4.1T}), in $n=2$ with $\theta_1$ matrix, simplifies to
\begin{equation}
\begin{split}
\Pi^{\mu\nu}(p)\big|_{n=2}^{\theta_1}&=\frac{e^2}{(4\pi)^2}[(g^{\mu\nu}p^2 - p^\mu p^\nu)\Pi_1
\\&+(g^{\mu\nu}(\theta p)^2-(\theta\theta)^{\mu\nu}p^2
+ p^{\{\mu}(\theta\theta p)^{\nu\}})\Pi_3
+(\theta p)^\mu (\theta p)^\nu(\Pi_2-\theta^2\Pi_4-2\theta^2\Pi_5 )],
\end{split}
\label{Pi12}
\end{equation}
where (for general $\theta$ matrix) the coefficients are given in (\ref{4.6}) and/or (\ref{4.11}), respectively.  \\

The UV part of (\ref{4.2}) obtained by using $\theta_1$ and/or (\ref{Pi12}) is then given below:
\begin{gather}
\begin{split}
\Pi^{\mu\nu}(p)\big|^{\theta_1}_{\rm UV}&=\frac{e^2}{48\pi^2}\bigg(\frac{2}{\epsilon}+
\ln\big(\mu^2 (\theta p)^2\big)\bigg)\bigg\{(g^{\mu\nu} p^2-p^\mu p^\nu)
\bigg[\big(3\kappa-1\big)^2-6\kappa^2\frac{p^2 \theta^2}{ (\theta p)^2}\bigg]
\\&
+\frac{1}{2}(g^{\mu\nu}(\theta p)^2-(\theta\theta)^{\mu\nu}p^2
+ p^{\{\mu}(\theta\theta p)^{\nu\}})\frac{ p^2}{(\theta p)^2}\bigg[11\kappa^2+2\kappa-1\bigg]
\\&
+\frac{(\theta p)^\mu (\theta p)^\nu}{(\theta p)^2} p^2\bigg[
73\kappa^2-86\kappa+25 + \big( \kappa-1\big)^2\frac{p^2 \theta^2}{ (\theta p)^2}\bigg]\bigg\}.
\end{split}
\label{PiUV1}
\end{gather}

A simple inspection of the above result shows that there is not any single $\kappa$ value which would simultaneously eliminate IR and UV plus log divergences, thus elimination of UV pathologies has to be treated more carefully. What is more, since in the UV regime we have $\Pi_i=B_i$ because of the absence of the tadpole contributions, the elimination of pathologies goes beyond choices of freedom parameter. However, a fine property is that UV part of polarization tensor (\ref{PiUV1}) in all limits $p\to 0, \; \theta \to 0$, separately and/or simultaneously, behave well enough. This is encouraging enough to consider that the UV plus log divergences in (\ref{PiUV1}) may be removed for any point $\kappa$ via certain proper subtraction of the linear combinations of dimensionless nonlocal counterterms, as noted in \cite{Horvat:2011bs},
\begin{equation}
{\cal B}_C=\xi\; \partial^2\frac{\tr\theta\theta}{(\theta \partial)^2}+\zeta\:\partial^2\frac{(\theta\theta \partial)^2}{(\theta \partial)^4},
\label{Counterterm}
\end{equation}
with $\xi,\zeta$ being free coefficients measuring strengths of the two terms. Those coefficients could be determined during the renormalization procedure, which we leave for the future as a part of the  NCGFT renormalization project.

\subsubsection{The subdimension  $n=4$ and $\theta^{\mu\nu}_2$ matrix}

Second choice is to set $n=4$ and selects the following $\theta^{\mu\nu}_2$ matrix\footnote{This condition was used in the renormalizability studies of four-dimensional NCGFT without the SW map \cite{Blaschke:2009aw,Blaschke:2010ck,Gomis:2000zz}. Note also that this $\theta^{\mu\nu}_2$ is full rank and, thus, breaks in general the unitarity if one performs Wick rotation to the Minkowski space\-time. It contains also time-space noncommutativity which breaks causality.}
 \cite{Horvat:2013rga}:
\begin{equation}
\theta^{\mu\nu}\equiv
\theta^{\mu\nu}_2=\theta^2
\begin{pmatrix}
0&-1&0&0\\
1&0&0&0\\
0&0&0&-1\\
0&0&1&0
\end{pmatrix}
=\theta^2
\begin{pmatrix}
{i\sigma_2}&0\\
0&{i\sigma_2}
\end{pmatrix}
\equiv(\sigma_2\otimes I_2)\;{\theta^2},
\label{nonudegen2}
\end{equation}
with $\sigma_2$ being famous Pauli matrix. This choice, which we are calling nonunitary, in four-dimensional Euclidean space\-time induces useful relations:
\begin{eqnarray}
(\theta\theta)^{\mu\nu}=- g^{\mu\nu}\theta^2,\,(\theta\theta p)^\mu=-p^\mu\theta^2, \,(\theta\theta\theta p)^{\mu}=-({\theta p})^\mu\theta^2,\,
(\theta p)^2\tr\theta\theta=-4(\theta\theta p)^2=-4\theta^4 p^2, \forall p; \;\theta^2=1/\Lambda_{\rm NC}^4.
\label{theta2}
\end{eqnarray}
The general tensor structure (\ref{4.1T}) then simplifies into two parts:
\begin{equation}
\begin{split}
\Pi^{\mu\nu}(p)|^{\theta_2}_{n=4}=\frac{e^2}{(4\pi)^2}\bigg[&\Big(g^{\mu\nu}p^2 - p^\mu p^\nu\Big)\Big(\Pi_1+2\frac{(\theta p)^2}{p^2}\Pi_3 -\frac{(\theta p)^4}{p^4}\Pi_4\Big)
+(\theta p)^\mu (\theta p)^\nu\Big(\Pi_2-2\frac{(\theta p)^2}{p^2}\Pi_5 \Big)\bigg].
\end{split}
\label{Pi24}
\end{equation}
The coefficients $\Pi_i$'s are given in (\ref{4.2}), (\ref{4.6}) and (\ref{4.11}), respectively.\\

The UV ($1/\epsilon$) divergent and the UV/IR mixing logarithmic terms in the total photon two-point function, come entirely from the bubble diagram, as one can easily be convinced by combining Eqs. (\ref{4.2}), (\ref{4.5UV}), and (\ref{Pi24}),
\begin{gather}
\begin{split}
\Pi^{\mu\nu}(p)|^{\theta_2}_{\rm UV}&=\frac{e^2 p^2}{24\pi^2}\bigg(\frac{2}{\epsilon}+
\ln\big(\mu^2 (\theta p)^2\big)\bigg)\bigg[\Big(g^{\mu\nu}-\frac{p^\mu p^\nu}{p^2}\Big)\big(3\kappa-1\big)^2
+\frac{(\theta p)^\mu (\theta p)^\nu}{(\theta p)^2}\frac{3}{2}\big(3\kappa-1\big)\big(9\kappa-7\big)\bigg].
\end{split}
\label{PiUV2}
\end{gather}
The above two divergences are certainly removable by selecting the special point $\kappa=1/3$~\cite{Horvat:2013rga,Trampetic:2014dea}. However, the UV plus log divergences from bubble diagrams are retained for any other $\kappa$ point in the $n=4$, $\theta_2$ case, and could be only removed by a proper subtraction of a linear combinations of nonlocal counterterms similar to (\ref{Counterterm}).

\subsubsection{Photon polarization for the $\theta_2$ matrix and $\kappa=1/3$ deformation}

In this case, because of the elimination of the UV divergences (\ref{PiUV2}), we analyze next the four-dimensional IR divergences
for $\theta^{\mu\nu}=\theta_{2}^{\mu\nu}$ satisfying $(\theta\theta)^{\mu\nu}=-g^{\mu\nu} \theta^2$ in four-dimensional Euclidean spacetime. So, the three-photon bubble plus four-photon-tadpole one-loop contributions to the photon two-point function, for special choice for $\theta_2$ and $\kappa=1/3$ is being reduced to two IR (UV/IR mixing) terms from photon tadpole diagram. This is obtained by using decomposition (\ref{Pi24}) with (\ref{Tik}) and (\ref{Tik'}), respectively. Adding two finite terms from the bubble diagram \cite{Horvat:2013rga},
\begin{equation}
\Pi_{\rm(I,II)}^{\mu\nu}(p)|^{\theta_2}_{\kappa=1/3}=\frac{e^2p^2}{2\pi^2}\bigg\{\frac{7}{27}\Big(g^{\mu\nu}-\frac{p^\mu p^\nu}{p^2}\Big)-\frac{1}{2}\frac{(\theta p)^\mu (\theta p)^\nu}{(\theta p)^2}\bigg\},
\label{4.16}
\end{equation}
we finally obtain the total photon polarization tensor in both models as
\begin{gather}
\begin{split}
\Pi_{\rm (I)}^{\mu\nu}(p)|^{\theta_2,\kappa=1/3}_{\rm IR}&=\frac{e^2p^2}{2\pi^2}\bigg\{\frac{7}{27}\Big(g^{\mu\nu}-\frac{p^\mu p^\nu}{p^2}\Big)\Big(1+\frac{4}{7}\frac{1}{p^2(\theta p)^{2}}\Big)
\\
&-\frac{1}{2}\frac{(\theta p)^\mu (\theta p)^\nu}{(\theta p)^2}\Big(1+\frac{4}{27}\frac{1}{p^2(\theta p)^2}\big(31-72\kappa_1-36\kappa_2+18\kappa_3-9\kappa_4\big)\Big)\bigg\},
\\
\Pi_{\rm (II)}^{\mu\nu}(p)|^{\theta_2,\kappa=1/3}_{\rm IR}&=\frac{e^2p^2}{2\pi^2}\bigg\{\frac{7}{27}\Big(g^{\mu\nu}-\frac{p^\mu p^\nu}{p^2}\Big)\Big(1+\frac{4}{7}\frac{1}{p^2(\theta p)^{2}}\Big)
\\
&-\frac{1}{2}\frac{(\theta p)^\mu (\theta p)^\nu}{(\theta p)^2}\Big(1+\frac{4}{27}\frac{1}{p^2(\theta p)^2}\big(22-36\kappa'_1-72\kappa'_2+18\kappa'_3\big)\Big)\bigg\}.
\end{split}
\label{4.18}
\end{gather}
The ``additional'' term $(\theta p)^{\mu}(\theta p)^{\nu}/(\theta p)^4$ receives multiple corrections from the tadpole diagram and can be easily removed by shifting the second-order gauge freedom parameters. Usual tensor structure $(g^{\mu\nu}p^2-p^{\mu}p^{\nu})$, however, depends solely on the first order gauge freedom parameter $\kappa$ in such a way that no real $\kappa$ value can set it to zero in models I and II.\\

\noindent
 At the end of this section let's summarize what we have learned about photon polarization tensor pathologies within models (I) and (II):\\
-- Summing over the tadpole and bubble diagram not only formally completes the one loop quantum corrections to the photon polarization tensor, but also appears to be crucial for the elimination of the quadratic IR divergences $\forall \theta$.\\
-- The quadratic IR divergences can be removed for arbitrary $\theta$, and for $\kappa=\kappa_1=\kappa_2=\kappa'_2=1$, plus appropriate choices for the rest of the freedom parameters in both models I and II.\\
-- Unfortunately, choice $\kappa=1$ does not remove UV divergences for arbitrary $\theta$.\\
-- Choice $\kappa=1/3$ and fixing $\theta=\theta_2$ removes UV but not IR for both models I and II.\\
-- We conclude that to eliminate simultaneously all divergences from tadpole plus bubble contributions requires a new extended model/procedure of deformation freedom parameter selections.

\section{Extending SW map-based models}

\subsection{Generalized SW map inspired four-photon self-interaction model}

Our experience in the last two sections suggests that in order to efficiently eliminate
pathological divergences in a deformed one-loop photon polarization tensor it is necessary
to modify the freedom parameter dependence at least in the four-photon interactions. Here we
propose such a modification to the SW mapped model I,\footnote{The same could be done also
with the model II action (\ref{2.16}), but we choose to discuss only one case for simplicity.}
which we call the model (III), to fulfill this requirement. We start by realizing
that $\kappa^2$ and $\kappa$ terms in (\ref{2.8}) and/or (\ref{2.10}) and/or (\ref{2.15}) may
be varied independently due to the manifestly gauge invariant structure of both terms,
so by this means we could assign independent gauge-symmetry (variation) freedom parameters to all five gauge invariant structures starting at $\theta^2$ order as found in \eqref{thetasquare}. Proceed this idea further we get the following action:
\begin{equation}
\begin{split}
S^{e^2}_{\eta_1,\eta_2,\eta_3,\eta_4,\eta_5}=&-\frac{e^2}{4}\theta^{ij}\theta^{kl}\int\,\eta_1(f_{\mu i}\star_2 f_{\nu j})(f^\mu_{\;\,\; k}\star_2 f^\nu_{\;\,\; l})-\eta_2(f_{ij}\star_2 f_{\mu\nu})(f^\mu_{\;\,\; k}\star_2 f^\nu_{\;\,\; l})
\\&+2\eta_3 f^{\mu\nu}[f_{\mu i}f_{\nu k}f_{jl}]_{\star_{3'}}-\frac{\eta_4}{4}f^{\mu\nu}\left[f_{\mu\nu}f_{ik}f_{jl}\right]_{\star_{3'}}
+\frac{\eta_5}{8}\left(f^{\mu\nu}\star_2 f_{ij}\right)\left(f_{kl}\star_2 f_{\mu\nu}\right)
\\&+2 f^{\mu\nu}(a_i\star_2\partial_j(f_{\mu k}\star_2 f_{\nu l})-[f_{\mu k} a_i\partial_j f_{\nu l}]_{\star_{3'}}-[a_i f_{\mu k}\partial_j f_{\nu l}]_{\star_{3'}})+\frac{1}{2}\theta^{pq}f^{\mu\nu}\left[\partial_i f_{jk} f_{lp}\partial_q f_{\mu\nu}\right]_{\mathcal M_{\rm (I)}}.
\end{split}
\label{7.1}
\end{equation}
Comparing \eqref{7.1} with \eqref{2.15} we see the following replacement rules: $\kappa^2:=\eta_1, \;\kappa:=\eta_2, \;\kappa_1:=\eta_3, \;\kappa_2:=1,\; \kappa_3:=\eta_4, \;\kappa_4:=\eta_5$. Here we omitted $\kappa_2$ because its correspond structure starts at $\theta^4$.

Interaction (\ref{7.1}) leads to the following result for the tadpole diagram in the $\eta$ setting
\begin{equation}
\begin{split}
T^{\mu\nu}(p)_{\eta}&=\frac{e^2}{(4\pi)^2}\{[g^{\mu\nu}p^2-p^\mu p^\nu]T_1(p)_{\eta}
+(\theta p)^\mu (\theta p)^\nu T_2(p)_{\eta}
\\&+[g^{\mu\nu}(\theta p)^2-(\theta\theta)^{\mu\nu}p^2
+ p^{\{\mu}(\theta\theta p)^{\nu\}}]T_3(p)_{\eta}
\\&+[(\theta\theta)^{\mu\nu}(\theta p)^2+(\theta\theta p)^\mu(\theta\theta p)^\nu]T_4(p)_{\eta}
+ (\theta p)^{\{\mu} (\theta\theta\theta p)^{\nu\}} T_5(p)_{\eta}\},
\end{split}
\label{4.1Teta}
\end{equation}
with $T_i(p)_{\eta}$ being
\begin{equation}
\begin{split}
T_1(p)_{\eta}&=\frac{4}{3}\left(\frac{tr\theta\theta}{(\theta p)^4}+4\frac{(\theta\theta p)^2}{(\theta p)^6}\right)\big(\eta_5-1\big),
\\
T_2(p)_{\eta}&=\frac{64}{3}\frac{1}{(\theta p)^4} \,\big(2\eta_1-4\eta_2+6\eta_3-2\eta_4+\eta_5+1\big),
\\
T_3(p)_{\eta}&=\frac{64}{3}\frac{1}{(\theta p)^4}\big(2\eta_1-2\eta_2+\eta_3+1\big),
\\
T_4(p)_{\eta}&=\frac{128}{3}\frac{p^2}{(\theta p)^6}\big(\eta_1-2\eta_2+\eta_3+1\big),
\\
T_5(p)_{\eta}&=\frac{64}{3}\frac{p^2}{(\theta p)^6}\big(2\eta_2-\eta_3-1\big).
\end{split}
\label{Tiketa}
\end{equation}

\subsection{Bubble plus tadpole results in model (III)}

We now consider the divergences from bubble plus tadpole diagrams in model III for both $\theta_1$ and $\theta_2$ choices. As we will see below, the quadratic IR divergence cancellation can be achieved for arbitrary $\kappa$ and both choices of $\theta$, while UV cancellation is only available for $\theta_2$.

\subsubsection{The IR part for subdimension $n=2$ and the $\theta_1$ matrix in the $\eta$ setting}

Substituting \eqref{Tiketa} into \eqref{4.5IR} then \eqref{Pi12}, we obtain the following conditions for quadratic IR divergence cancellation
\begin{equation}
\begin{split}
2(1-\kappa)^2-(\eta_5-1)=0,
\\
\big(1-3\kappa\big)\big(3-\kappa\big)-2(2\eta_1-4\eta_2+6\eta_3-2\eta_4+\eta_5+1)=0,
\\
(1-10\kappa+17\kappa^2)-4(2\eta_1-2\eta_2+\eta_3+1)=0,
\\
(3\kappa^2-2\kappa+1)-2\eta_1=0,
\end{split}
\end{equation}
with solutions
\begin{equation}
\begin{split}
\eta_1=&\frac{1}{2}\big(3\kappa^2-2\kappa+1\big),\;
\eta_3=2\eta_2+\frac{1}{4}\big(5\kappa^2-2\kappa-7\big),
\\
\eta_4=&4\eta_2-\frac{1}{2}\big(11\kappa^2-4\kappa-7\big),\;
\eta_5=1+2\big(\kappa-1\big)^2,
\end{split}
\label{7.9}
\end{equation}
which produce the IR free photon polarization tensor for arbitrary $\kappa$, as expected.

\subsubsection{Subdimension $n=4$ represented by the $\theta_2$ matrix and for the $\eta$ setting}

For NCQFTs in $d$ dimensions, equal to the subspace dimensions $n$, i.e. in four-dimensional Euclidian spacetime with
$n=4$ case, and by choosing $\theta_2$ deformation (\ref{nonudegen2}), we get the following tadpole
\begin{equation}
T^{\mu\nu}(p)_{\eta}|^{n=4,\theta_2}_{\rm IR}=\frac{e^2}{3\pi^2}\frac{1}{(\theta p)^2}\bigg[\Big(g^{\mu\nu} - \frac{p^\mu p^\nu}{p^2}\Big)\,2\eta_1
+\frac{(\theta p)^\mu (\theta p)^\nu}{(\theta p)^2}\big(2\eta_1-8\eta_2+8\eta_3-2\eta_4+\eta_5+3\big)\bigg].
\label{7.4}
\end{equation}

Using the above Eq. (\ref{7.4}), together with (\ref{4.1T}) and (\ref{4.2}) for $\theta_{2}$,
we obtain the following bubble plus tadpole contributions to the photon polarization tensor in the IR regime:
\begin{gather}
\begin{split}
\Pi^{\mu\nu}(p)|^{\theta_2,\kappa,\eta_i}_{\rm IR}=&\frac{e^2}{6\pi^2}\frac{1}{(\theta p)^{2}}\bigg[\Big(g^{\mu\nu}-\frac{p^\mu p^\nu}{p^2}\Big)\Big(-3\kappa^2-2\kappa+1+4\eta_1\Big)
\\
&-\frac{(\theta p)^\mu (\theta p)^\nu}{(\theta p)^2}\Big(15\kappa^2-26\kappa+7-4\eta_1+16\eta_2-16\eta_3+4\eta_4-2\eta_5-6\Big)\bigg].
\end{split}
\label{4.17}
\end{gather}
Here, due to the $\eta$ settings the quadratic IR tadpole contributions are clearly disentangled from the well-behaving rest of contributions. Since the number of free parameters exceeds the number of tensor structures in a subspace $n=4$, the quadratic IR divergence cancellation surely exists for arbitrary $\kappa$.
Furthermore, an interesting choice,
\begin{equation}
\kappa=\frac{1}{3},\; \eta_1=0,\; \eta_2=\eta_3, \;2\eta_4-\eta_5=3,
\label{ketai3}
\end{equation}
eliminates not only IR but, simultaneously due to (\ref{PiUV2}), also the hard UV and logarithmic divergences as well.

\section{Discussion}

Following recent progress in the $e^3$-order $\theta$-exact Seiberg-Witten map expansions \cite{Martin:2012aw,Trampetic:2015zma,Martin:2015nna},  in this paper the resulting four-photon interaction is presented through the construction of three different models (I,II,III). We study their impact on the perturbative quantization of the SW map deformed $\rm U(1)$ gauge field theory via the corresponding one-loop, four-photon-tadpole contributions to the (divergent part of the) photon polarization tensor. Note also that the same term should contribute to the NC phenomenology at extreme energies, for example tree-level NCQED contributions to the $\gamma\gamma\to\gamma\gamma$ scattering processes~\cite{Hewett:2000zp}.

Quadratic IR divergence in NCQFT on Moyal space was first found in the original version of UV/IR mixing in the tadpole integral of the NC $\phi^4$ QFT on Moyal space~\cite{Matusis:2000jf}, where part of the originally quadratically UV divergent diagram gets UV regularized from the NC phase factor and becomes quadratically IR divergent instead. The UV divergence in the commutative $\rm U(1)$ gauge theory, on the other hand, is logarithmic under the dimensional regularization procedure, yet the IR divergence (steaming from UV/IR mixing) in the NC $\rm U_\star(1)$ gauge theory still starts from the quadratic order~\cite{Hayakawa:1999yt,Hayakawa:1999zf,Minwalla:1999px}. It is shown in our previous work on the bubble diagram~\cite{Horvat:2013rga} with the first-order gauge freedom parameter $\kappa_g$ $(=\kappa^{-1})$ that the hard UV ($1/\epsilon$) divergence, the commutative logarithmic divergence $\ln (\mu^2/p^2)$ and the NC logarithmic divergence $\ln p^2(\theta p)^2$, share the same dependence on $\kappa_g$, while the quadratic IR divergence bears a completely different one. Thus the quadratic IR divergence may indeed have a separated origin from ``UV/IR mixing'' which is logarithmic as it should be.

By evaluating the photon one-loop tadpole diagram Fig.\ref{fig:photontadpole} in the first two deformed $\rm U(1)$ gauge theory models based on two distinct gauge field strengths~\cite{Trampetic:2015zma} using NC-extended dimensional regularization techniques, we show that the NC massless tadpole integrals produce solely quadratic IR divergence, i.e. there is no UV divergent or finite contributions from tadpole. From this perspective the he result of this paper suggests that the quadratic IR divergence behavior could be simply a tadpole effect, especially because it can be removed by a suitable gauge freedom selection. The two models give different tadpole contributions for general values of the SW / gauge freedom deformation parameters $\kappa$, $\kappa_i$'s and $\kappa'_i$'s, yet they can be made equal by employing certain simple algebraic relations. In particular the tadpole integrals from the two models are equal to each other for $\kappa_i=\kappa'_i=1, \forall i$. In fact this is not surprising as the limited number of momenta involved in the tadpole evaluation restrict the possible value for the nonlocal factors to either be unity or $\sin^2\frac{p\theta k}{2}/\big(\frac{p\theta k}{2}\big)^2$, thus severely constraining the possible combinations in the tensor structure(s).

After summing of bubble and tadpole diagrams in the above three models, we study the possible choices of freedom parameters for divergence eliminations. These choices are expectedly different in the so-called $\kappa$-setting models (I, II) and in the third $\eta$-setting model (III). Of importance is that $\kappa=1$ stands out as the unique quadratic IR divergence cancelation point for arbitrary $\theta$ and for both models (I, II), while UV and log/mixing divergence can only be removed at a different $\kappa=\frac{1}{3}$ point with the special choice for  $\theta$, $\theta=\theta_2$ matrix. For the third $\eta$-setting model the quadratic IR divergence removal is available for arbitrary $\kappa$ and both $\theta_{1,2}$ choices, as expected. We, thus, manage to obtain the full divergence cancellation at the $(\kappa,\; \eta_1)=(\frac{1}{3},\;0)$, point for $\theta=\theta_2$ matrix with the help of extra freedom parameters $\eta_2=\eta_3$, $2\eta_4-\eta_5=3$ in the four-photon interaction (\ref{ketai3}).

From the $\theta$-dependent photon polarization tensor in the U(1) NCQFT with
the special choice for $\theta_2$, we raise one more interesting question: 
Does our deformation parameter set $(\kappa,\eta_i)$ run between IR and UV divergence-eliminating points? 
To be more precise, let us take a liberty of assuming that there is indeed a possibility
that deformation parameters $\kappa$ and $\eta$ are energy/momentum-dependent
quantities.\footnote{That would not be strange at all, since our deformation parameters
are actually quantities similar to coupling constants---sitting in the actions---of any QFT.} As
a next point after inspecting (\ref{PiUV2}), (\ref{4.17}) and (\ref{ketai3}) we notice the fact that starting
with the photon polarization behavior in the deep IR regime, and moving towards
the high UV regime, our divergence eliminating freedom parameters  $\kappa,\eta_i$
decreases: $\kappa=1\,\to\,\frac{1}{3},\; \eta_1=1\,\to\,0, \;2\eta_4-\eta_5=7\,\to\,3$, while $\eta_2-\eta_3=0$. Now taking into account the results of the $\theta^1$ models \cite{Buric:2006wm,Latas:2007eu}, where in the NC SU(N) model \cite{Latas:2007eu} was explicitly shown that $\theta^1$ parameter runs while having a negative $\beta$ function, we conjecture that $\beta$ functions associated with the parameters $\kappa$, $\eta_1$, and $(2\eta_4-\eta_5)$ could be negative too.  The $(\eta_2-\eta_3)$ difference should have a zero $\beta$ function. However a precise computation of  the above $\beta$ functions goes well beyond the scope of this paper and, therefore, such considerations should be left for the future.

\section{Conclusion}

We introduce three different types of nonlocal four-photon interactions in a full-fledged $\theta$-exact deformed U(1) gauge field theory. The first two actions (I,II) are deformed by the two distinct $\theta$-exact gauge field strength Seiberg-Witten (SW) maps, and the third one (III) involves a SW-map inspired gauge invariance deformation in a more general way.

The four-photon interaction induces the tadpole diagram which results in a pure quadratic IR divergence. Based on the computation of that tadpole we conjecture that the general origin of the quadratic IR divergence, including those from the bubble diagram, stems from the tadpolelike integrals, which although highly UV divergent still vanish under the dimensional regularization in the commutative theory~\cite{Leibbrandt:1975dj}.

In conclusion, by performing the sum over one-loop bubble and tadpole diagrams in all
three models, we obtain for arbitrary noncommutative tensors $\theta^{\mu\nu}$ and with
suitable  choices of freedom parameters, the quadratic IR divergence free-photon
polarization tensor. A simultaneous UV, quadratic IR, and logarithmic divergences
cancellation does occur only in model III by employing the special $\theta=\theta_2$
matrix and the deformation parameter set,  $\kappa=\frac{1}{3},\; \eta_1=0,\; \eta_2=\eta_3, \;2\eta_4-\eta_5=3$. The most important fact is definitely a huge freedom due to the SW mapping and gauge invariance, which at the end of the day allows elimination of all pathologies in the one-loop two point function, which is quite aspiring although there is  still a very long way before one could speculate a renormalizable U(1) theory on Moyal space. The IR improvement we have achieved till now seems to suggest that there should be some way 
out within the SW map approach, therefore we believe that further investigation along this 
line could help reaching an accurate formulation with SW map and 
provide  some answers about the renormalizability of the NCQFT in general.

Besides the splitting of the polarization states driven by the breaking of Lorentz invariance caused by $\theta^{\mu \nu}$, the persistence of a finite additional term/effect (\ref{4.16}) of quantum-gravity origin may also affect the low-momentum end of the spectrum in the photon dispersion relation. The serious possibility of eliminating all pathological effects in the photon polarization tensor within NCQFT is our main motivation to continue searching for the noncommutative-geometry/quantum-gravity-inspired UV/IR mixing phenomena and connections between the NCQFT and holography, black hole horizon physics, etc. \cite{Horvat:2010km,Horvat:2011bs,Horvat:2011qg,Aschieri:2012in,Horvat:2010sr}.

\section{Acknowledgments}
The work by R.H. and J.Y. has been fully supported by Croatian Science Foundation under Project No. IP-2014-09-9582. The work  J.T. is conducted under the European Commission and the Croatian Ministry of Science, Education and Sports Co-Financing Agreement No. 291823. In particular, J.T. acknowledges project financing by the Marie Curie FP7-PEOPLE-2011-COFUND program NEWFELPRO: Grant Agreement No. 69.
J.T. would also like to acknowledge the partial support of the Alexander von Humboldt Foundation (KRO 1028995 STP), and  Max-Planck-Institute for Physics, and W. Hollik for hospitality. 
J.Y. would like to acknowledge the Alexander von Humboldt Foundation and COST Action MP1405 for partially supporting his participation in the Corfu Summer Institute 2015 as well as the organizers of the Corfu Summer Institute 2015 for hospitality.
We would like to thank P. Aschieri, D. Blaschke, M. Dimitrijevi\'c \'Ciri\'c, J. Erdmenger, M. Hanada,  W. Hollik, A.Ilakovac, T. Juri\'c, C. P. Martin, P. Schupp, and R. Szabo for fruitful discussions. A great deal of computation was done by using MATHEMATICA 8.0Mathematica \cite{mathematica} plus the tensor algebra package xACT~\cite{xAct}. Special thanks to A. Ilakovac and D. Kekez for the computer software and hardware support.

\appendix

\section{Explicit gauge invariant action}

The $\rm U(1)$ gauge invariance of the four-photon interaction \eqref{2.8} and \eqref{2.9} is quite lengthy to verify by directly evaluating the infinitesimal gauge transformation. It is convenient to put them into the same form as \eqref{3photon} but not an easy task because of the loss of cyclicity. In this appendix we show that this step is achievable by analyzing the integration-by-part in the momentum space.

We first notice that the generalized star products based on the constant Moyal deformation parameter $\theta^{ij}$ bear a relatively common form in momentum space
\begin{equation}
\left[f_1...f_n\right]_{\mathcal M}\left(x\right)=\int\prod\limits_{i=1}^n \frac{d^dp_i}{(2\pi)^d} \prod\limits_{i=1}^n\tilde f_i(p_i)\exp\bigg[-i\Big(\sum\limits_{i=1}^n p_i\Big)x\bigg]{\cal F}\left(p_1,...,p_n;\theta^{ij}\right).
\end{equation}
A few examples relevant to this paper include
\begin{gather}
\begin{split}
(f\star g)(x)&=\int\,e^{-i(p+q)x}\tilde f(p)\tilde g(q)f_{\star}\left(p,q\right),\;\;
(f\star_2 g)(x)=\int\,e^{-i(p+q)x}\tilde f(p)\tilde g(q)f_{\star_2}\left(p,q\right),
\\
[f g h]_{\star_3}(x)&=\int\,e^{-i(p+q+k)x}\tilde f(p)\tilde g(q)\tilde h(k)f_{\star_3}\left(p,q,k\right),\;\;
[f g h]_{\star_{3'}}(x)=\int\,e^{-i(p+q+k)x}\tilde f(p)\tilde g(q)\tilde h(k)f_{\star_{3'}}\left(p,q,k\right),
\label{2.7}
\end{split}
\end{gather}
with
\begin{gather}
\begin{split}
f_{\star}(p,q)&=\exp\Big(i\frac{p\theta q}{2}\Big),\;\;f_{\star_2}(p,q)=\frac{\sin\frac{p\theta q}{2}}{\frac{p\theta q}{2}},
\\
f_{\star_{3}}\left(p,q,k\right)&=\frac{\sin \frac{p\theta k}{2}\sin(\frac{p\theta q}{2}+\frac{p\theta k}{2})}{(\frac{p\theta q}{2}+\frac{p\theta k}{2})(\frac{p\theta k}{2}+\frac{q\theta k}{2})}+\frac{\sin \frac{q\theta k}{2}\sin(\frac{p\theta q}{2}-\frac{q\theta k}{2})}{(\frac{p\theta q}{2}-\frac{q\theta k}{2})(\frac{p\theta k}{2}+\frac{q\theta k}{2})},
\\
f_{\star_{3'}}\left(p,q,k\right)&=\frac{\cos(\frac{p\theta q}{2}+\frac{p\theta k}{2}-\frac{q\theta k}{2})-1}{(\frac{p\theta q}{2}+\frac{p\theta k}{2}-\frac{q\theta k}{2})\frac{q\theta k}{2}}-\frac{\cos(\frac{p\theta q}{2}+\frac{p\theta k}{2}+\frac{q\theta k}{2})-1}{(\frac{p\theta q}{2}+\frac{p\theta k}{2}+\frac{q\theta k}{2})\frac{q\theta k}{2}}.
\end{split}
\label{A8}
\end{gather}

Since ${\cal F}\left(p_1,...,p_n;\theta^{ij}\right)$ is coordinate independent, one can introduce general integration-by-part transformation
\begin{equation}
\int d^d x g(x) \left[f_1...f_n\right]_{{\cal F}'}\left(x\right)=\int d^dx f_i(x) \left[f_1...f_{i-1} g f_{i+1}...f_n\right]_{{\cal F}}(x),
\end{equation}
by setting
\begin{equation}
{\cal F}\left(p_1,...p_{i-1},q_g,p_{i+1},...,p_n;\theta^{ij}\right)|_{q_g=-\sum\limits_{i=1}^n p_i}={\cal F}'\left(p_1,...,p_n;\theta^{ij}\right),
\end{equation}
because
\begin{equation}
\begin{split}
&\int d^dx f_i(x) \left[f_1...f_{i-1} g f_{i+1}...f_n\right]_{{\cal F}}(x)
\\&=\int d^d x\int\frac{d^d q_g}{(2\pi)^d}\prod\limits_{i=1}^n \frac{d^dp_i}{(2\pi)^d} \tilde g(q_g)\prod\limits_{i=1}^n\tilde f_i(p_i) \exp\bigg[-i\Big(q_g+\sum\limits_{i=1}^n p_i\Big)x\bigg]{\cal F}\left(p_1,...p_{i-1},q_g,p_{i+1},...,p_n;\theta^{ij}\right)
\\&=\int\frac{d^d q_g}{(2\pi)^d}\prod\limits_{i=1}^n \frac{d^dp_i}{(2\pi)^d} \tilde g(q_g)\prod\limits_{i=1}^n\tilde f_i(p_i) {\cal F}\left(p_1,...p_{i-1},q_g,p_{i+1},...,p_n;\theta^{ij}\right)(2\pi)^d\delta^d\Big(q_g+\sum\limits_{i=1}^n p_i\Big)
\\&=\int\prod\limits_{i=1}^n \frac{d^dp_i}{(2\pi)^d} \prod\limits_{i=1}^n\tilde f_i(p_i) \Big(\tilde g(q_g){\cal F}\left(p_1,...p_{i-1},q_g,p_{i+1},...,p_n;\theta^{ij}\right)\Big)\Big|_{q_g=-\sum\limits_{i=1}^n p_i}
\\&=\int\prod\limits_{i=1}^n \frac{d^dp_i}{(2\pi)^d} \prod\limits_{i=1}^n\tilde f_i(p_i) \tilde g\Big(q_g=-\sum\limits_{i=1}^n p_i\Big){\cal F}'\left(p_1,...,p_n;\theta^{ij}\right)
\\&=\int\frac{d^d q_g}{(2\pi)^d}\prod\limits_{i=1}^n \frac{d^dp_i}{(2\pi)^d} \tilde g(q_g)\prod\limits_{i=1}^n\tilde f_i(p_i){\cal F}'\left(p_1,...,p_n;\theta^{ij}\right)(2\pi)^d\delta^d\Big(q_g+\sum\limits_{i=1}^n p_i\Big)
\\&=\int d^d x g(x) \left[f_1...f_n\right]_{{\cal F}'}\left(x\right).
\end{split}
\end{equation}
Once ${\cal F}'$ equals ${\cal F}$ we obtain certain level of cyclicity in a slightly generalized sense. For example,
\begin{equation}
f_{\star}(p_3,p_1)|_{p_1=-(p_2+p_3)}=f_{\star}(p_2,p_3),
\end{equation}
gives the 3-cyclicity of the Moyal star product
\begin{equation}
\int f(x)(g(x)\star h(x))=\int g(x)(h(x)\star f(x)).
\end{equation}
Using the same method one can also show the following relations
\begin{equation}
\int f(g\star_2 h)=\int g (h\star_2 f)=\int h (g\star_2 f),\;\;
\int f(g\star_2(h\star_2 k))=\int g(f\star_2(h\star_2 k)),\;\;
\int f[g h k]_{\star_{3'}}=g[f h k]_{\star_{3'}}.
\end{equation}
The simple relations above are not yet sufficient to perform the integration-by-part that we want. However, they do indicate that one should look for more momentum space identities first. We achieve this goal by analyzing the Ward identity $p_{i_{\mu_i}}\Gamma^{...\mu_i...}, i=1,....,4$. Here we consider only the part of the action(s) which is not yet explicitly gauge invariant, for example from \eqref{2.8},
\begin{equation}
\begin{split}
S^{e^2}_{\rm (I)_r}=-\frac{e^2}{4}\theta^{ij}\theta^{kl}\int\,&(a_i\star_2\partial_j f_{\mu\nu})(a_k\star_2\partial_l f^{\mu\nu})
+\frac{1}{2}f^{\mu\nu}(2[a_i\partial_j a_k \partial_l f_{\mu\nu}]_{\star_{3'}}+2[\partial_l f_{\mu\nu}a_i \partial_j a_k]_{\star_{3'}}
\\&+2[a_i a_k \partial_j\partial_l f_{\mu\nu}]_{\star_{3'}}
-[\partial_l f_{\mu\nu}a_i\partial_k a_j]_{\star_{3'}}-[a_i\partial_k a_j\partial_l f_{\mu\nu}]_{\star_{3'}}).
\end{split}
\label{s1r}
\end{equation}
From the interaction above we can then read out a vertex
\begin{equation}
\begin{split}
\Gamma^{\mu_1\mu_2\mu_3\mu_4}_{\rm (I)_r}\propto& (2((p_1 p_4)g^{\mu_1\mu_4}-p_1^{\mu_4}p_4^{\mu_1})((2(\theta p_3)^{\mu_2}(\theta p_4)^{\mu_3}+2(\theta p_4)^{\mu_2}(\theta p_4)^{\mu_3}-(p_3\theta p_4)\theta^{\mu_2\mu_3})f_{\star_{3'}}[p_4,p_2,p_3]
\\&+(2(\theta p_3)^{\mu_2}(\theta p_4)^{\mu_3}-(p_3\theta p_4)\theta^{\mu_2\mu_3})f_{\star_{3'}}[p_4,p_2,p_3]+2(\theta p_1)^{\mu_2}(\theta p_4)^{\mu_3}f_{\star_2}(p_1,p_2)f_{\star_2}(p_3,p_4))
\\&+{\rm all\; S_4\; permutations\; over}\; \{p_i\}{\rm \; and}\; \{\mu_i\}{\rm \; simutaneously})\cdot(p_1+p_2+p_3+p_4).
\end{split}
\end{equation}
Next we find that the Ward identity is satisfied once we sum over the following four permutations: identity, $2\leftrightarrow 3$, $1\leftrightarrow 4$, and the composition of the last two. The rest of the $\rm S_4$ permutations only change the labeling for the contraction. Two identities emerge in this analysis:
\begin{gather}
f_{\star_{3'}}[p_3,p_3,p_4]+f_{\star_{3'}}[p_3,p_2,-(p_2+p_3+p_4)]=2f_{\star_2}(p_2,p_3+p_4)f_{\star_2}(p_3,p_4),
\\
\begin{split}
2(p_2\theta p_3)f_{\star_{3'}}[p_4,p_2,p_3]=&(p_3\theta p_4) f_{\star_{3'}}[p_2,p_3,p_4]-(p_3\theta (p_2+p_4)) f_{\star_{3'}}[p_2,p_3,-(p_2+p_3+p_4)]
\\&-(p_2\theta p_4) f_{\star_{3'}}[p_3,p_2,p_4]+(p_2\theta(p_3+p_4))f_{\star_{3'}}[p_3,p_2,p_1].
\end{split}
\end{gather}
We can construct the following integration-by-part relations out of these two identities
\begin{equation}
2\int f(g\star_2(h\star_2 k))=\int f[ghk]_{\star_{3'}}+k[hgf]_{\star_{3'}},
\label{IID1}
\end{equation}
and
\begin{equation}
2\theta^{ij}\int f[k\partial_i g\partial_j h]_{\star_{3'}}=\theta^{ij}\int f[g\partial_i h\partial_j k]_{\star_{3'}}+k[g\partial_i h\partial_j f]_{\star_{3'}}-f[h\partial_i g\partial_j k]_{\star_{3'}}-k[h\partial_i g\partial_j f]_{\star_{3'}}.
\label{IID2}
\end{equation}
Now we return to \eqref{s1r} and rearrange the first term using \eqref{IID1} and the last term using \eqref{IID2}, getting
\begin{gather}
\begin{split}
\int\,\theta^{ij}\theta^{kl}(a_i\star_2\partial_j f_{\mu\nu})(a_k\star_2\partial_l f^{\mu\nu})=&\frac{1}{2}\int\,\theta^{ij}\theta^{kl}\left(\partial_j f_{\mu\nu}[a_i a_k \partial_l f^{\mu\nu}]_{\star_{3'}}+\partial_l f_{\mu\nu}[a_k a_i \partial_j f^{\mu\nu}]_{\star_{3'}}\right)
\\
=&\int\,\theta^{ij}\theta^{kl}\partial_j f_{\mu\nu}[a_i a_k \partial_l f^{\mu\nu}]_{\star_{3'}},
\end{split}
\\
\begin{split}
\int\,\theta^{ij}\theta^{kl}f_{\mu\nu}[a_i\partial_k a_j\partial_l f{\mu\nu}]_{\star_{3'}}=&\frac{1}{4}\int\,\theta^{ij}\theta^{kl}f_{\mu\nu}([a_i\partial_k a_j\partial_l f^{\mu\nu}]_{\star_{3'}}+[a_i\partial_k a_j\partial_l f^{\mu\nu}]_{\star_{3'}}
\\&-[a_j\partial_k a_i\partial_l f^{\mu\nu}]_{\star_{3'}}-[a_j\partial_k a_i\partial_l f^{\mu\nu}]_{\star_{3'}})
\\=&\frac{1}{2}\int\,\theta^{ij}\theta^{kl}f^{\mu\nu}[f_{\mu\nu}\partial_k a_i\partial_l a_j]_{\star_{3'}}.
\end{split}
\end{gather}
One can also use the partial 4-cyclicity of $\star_{3'}$ to obtain
\begin{gather}
\int\,\theta^{ij}\theta^{kl}f^{\mu\nu}[\partial_l f_{\mu\nu}a_i\partial_k a_j]_{\star_{3'}}=\int\,\theta^{ij}\theta^{kl}\partial_l f_{\mu\nu}[f{\mu\nu}a_i\partial_k a_j]_{\star_{3'}}=-\frac{1}{2}\int\,\theta^{ij}\theta^{kl} f^{\mu\nu}[f_{\mu\nu}\partial_l a_i\partial_k a_j]_{\star_{3'}},
\\
\int\,\theta^{ij}\theta^{kl}f^{\mu\nu}[\partial_l f_{\mu\nu}a_i \partial_j a_k]_{\star_{3'}}=\int\,\theta^{ij}\theta^{kl}\partial_l f^{\mu\nu}[ f_{\mu\nu}a_i \partial_j a_k]_{\star_{3'}}=-\frac{1}{2}\int\,\theta^{ij}\theta^{kl} f^{\mu\nu}[ f_{\mu\nu}\partial_l (a_i \partial_j a_k)]_{\star_{3'}}.
\end{gather}
Now \eqref{s1r} boils down to
\begin{equation}
\begin{split}
S^{e^2}_{\rm (I)_r}=&-\frac{e^2}{4}\theta^{ij}\theta^{kl}\int\,-f^{\mu\nu}\left[\partial_j a_i a_k\partial_l f_{\mu\nu}\right]_{\star_{3'}}-\frac{1}{2}f^{\mu\nu}[ f_{\mu\nu}a_k \partial_l \partial_j a_i]_{\star_{3'}}+\frac{1}{2}f^{\mu\nu}(\left[f_{\mu\nu}\partial_k a_i\partial_l a_j\right]_{\star_{3'}}-\left[f_{\mu\nu}\partial_l a_i\partial_j a_k\right]_{\star_{3'}})
\\=&-\frac{e^2}{4}\theta^{ij}\theta^{kl}\int\,-f^{\mu\nu}\left[\partial_j a_i a_k\partial_l f_{\mu\nu}\right]_{\star_{3'}}-\frac{1}{2}f^{\mu\nu}[ f_{\mu\nu}a_k \partial_l \partial_j a_i]_{\star_{3'}}-\frac{1}{4}f^{\mu\nu}\left[f_{\mu\nu}f_{ik}f_{jl}\right]_{\star_{3'}}.
\end{split}
\end{equation}
We then notice that the above first term is
\begin{equation}
\begin{split}
\int\,\theta^{ij}\theta^{kl}f^{\mu\nu}\left[\partial_j a_i a_k\partial_l f_{\mu\nu}\right]_{\star_{3'}}=&-\int\,\theta^{ij}\theta^{kl}(\partial_l f^{\mu\nu}\left[\partial_j a_i a_k f_{\mu\nu}\right]_{\star_{3'}}+f^{\mu\nu}\left[\partial_j a_i \partial_l a_k f_{\mu\nu}\right]_{\star_{3'}}+f^{\mu\nu}\left[\partial_l\partial_j a_i a_k f_{\mu\nu}\right]_{\star_{3'}})
\\=&-\int\,\theta^{ij}\theta^{kl}(\partial_l f^{\mu\nu}\left[\partial_j a_i a_k f_{\mu\nu}\right]_{\star_{3'}}+\frac{1}{4}f^{\mu\nu}\left[f_{ij}f_{kl} f_{\mu\nu}\right]_{\star_{3'}}+f^{\mu\nu}\left[\partial_l\partial_j a_i a_k f_{\mu\nu}\right]_{\star_{3'}}),
\end{split}
\end{equation}
therefore
\begin{equation}
\begin{split}
S^{e^2}_{\rm (I)_r}=&-\frac{e^2}{4}\theta^{ij}\theta^{kl}\int\,\frac{1}{2}(\partial_l f^{\mu\nu}\left[\partial_j a_i a_k f_{\mu\nu}\right]_{\star_{3'}}-f^{\mu\nu}\left[\partial_j a_i a_k\partial_l f_{\mu\nu}\right]_{\star_{3'}})+\frac{1}{2}f^{\mu\nu}(\left[\partial_l\partial_j a_i a_k f_{\mu\nu}\right]_{\star_{3'}}
\\&-[ f_{\mu\nu}a_k \partial_l \partial_j a_i]_{\star_{3'}})
-\frac{1}{4}\left[f_{\mu\nu}f_{ik}f_{jl}\right]_{\star_{3'}}+\frac{1}{8}\left[f_{ij}f_{kl} f_{\mu\nu}\right]_{\star_{3'}}.
\end{split}
\end{equation}
The first two parentheses can be put into a common formula,
\begin{equation}
\begin{split}
&\theta^{ij}\theta^{kl}\int\,\frac{1}{2}(\partial_l f^{\mu\nu}\left[\partial_j a_i a_k f_{\mu\nu}\right]_{\star_{3'}}-f^{\mu\nu}\left[\partial_j a_i a_k\partial_l f_{\mu\nu}\right]_{\star_{3'}})
+\frac{1}{2}f^{\mu\nu}(\left[\partial_l\partial_j a_i a_k f_{\mu\nu}\right]_{\star_{3'}}-[ f_{\mu\nu}a_k \partial_l \partial_j a_i]_{\star_{3'}})
\\&=\frac{1}{4}\theta^{ij}\theta^{kl}\theta^{pq}\int\,f^{\mu\nu}\left[\partial_p\partial_j a_i \partial_q a_k\partial_l f_{\mu\nu}\right]_{\mathcal M_{\rm (I)}}-\partial_l f^{\mu\nu}\left[\partial_p\partial_j a_i \partial_q a_k f_{\mu\nu}\right]_{\mathcal M_{\rm (I)}}
\\&+f^{\mu\nu}\left[\partial_l\partial_j a_i \partial_p a_k\partial_q f_{\mu\nu}\right]_{\mathcal M_{\rm (I)}}-\partial_q f^{\mu\nu}\left[\partial_l\partial_j a_i \partial_p a_k f_{\mu\nu}\right]_{\mathcal M_{\rm (I)}}
\\&=\frac{1}{4}\theta^{ij}\theta^{kl}\theta^{pq}\int\,-f^{\mu\nu}\left[\partial_l\partial_j a_i \partial_k a_p\partial_q f_{\mu\nu}\right]_{\mathcal M_{\rm (I)}}+\partial_q f^{\mu\nu}\left[\partial_l\partial_j a_i \partial_k a_p f_{\mu\nu}\right]_{\mathcal M_{\rm (I)}}
\\&+f^{\mu\nu}\left[\partial_l\partial_j a_i \partial_p a_k\partial_q f_{\mu\nu}\right]_{\mathcal M_{\rm (I)}}-\partial_q f^{\mu\nu}\left[\partial_l\partial_j a_i \partial_p a_k f_{\mu\nu}\right]_{\mathcal M_{\rm (I)}}
\\&=\frac{1}{4}\theta^{ij}\theta^{kl}\theta^{pq}\int\,f^{\mu\nu}\left[\partial_i f_{jk} f_{lp}\partial_q f_{\mu\nu}\right]_{\mathcal M_{\rm (I)}}-\partial_q f^{\mu\nu}\left[\partial_i f_{jk} f_{lp} f_{\mu\nu}\right]_{\mathcal M_{\rm (I)}},
\end{split}
\end{equation}
by using a new 3-product
\begin{equation}
[fgh]_{\mathcal M_{\rm (I)}}(x)=\int\,e^{-i(p+q+k)x}\tilde f(p)\tilde g(q)\tilde h(k)f_{\rm (I)}\left(p,q,k\right),
\label{2.12}
\end{equation}
where
\begin{equation}
\begin{split}
f_{\rm (I)}\left(p,q,k\right)&=(p\theta q)^{-1}\left(f_{\star_{3'}}[p,q,-(p+q+k)]-f_{\star_{3'}}[p,q,k]\right)
\\&=\frac{2}{(\frac{p\theta q}{2}-\frac{p\theta k}{2}-\frac{q\theta k}{2})(\frac{p\theta q}{2}+\frac{p\theta k}{2}-\frac{q\theta k}{2})(\frac{p\theta q}{2}+\frac{p\theta k}{2}+\frac{q\theta k}{2})}
\\&+\frac{\cos(\frac{p\theta q}{2}-\frac{p\theta k}{2}-\frac{q\theta k}{2})}{2\frac{p\theta q}{2}(\frac{p\theta q}{2}-\frac{q\theta k}{2})(\frac{p\theta q}{2}-\frac{p\theta k}{2}-\frac{q\theta k}{2})}
+\frac{\cos(\frac{p\theta q}{2}+\frac{p\theta k}{2}-\frac{q\theta k}{2})}{2\frac{q\theta k}{2}(\frac{p\theta q}{2}-\frac{q\theta k}{2})(\frac{p\theta q}{2}+\frac{p\theta k}{2}-\frac{q\theta k}{2})}
\\&+\frac{\cos(\frac{p\theta q}{2}+\frac{p\theta k}{2}+\frac{q\theta k}{2})}{2\frac{p\theta q}{2}\frac{q\theta k}{2}(\frac{p\theta q}{2}+\frac{p\theta k}{2}+\frac{q\theta k}{2})}.
\end{split}
\end{equation}
It is not hard to show that the $\mathcal M_{\rm (I)}$ product satisfies an ``anticyclicity'',
\begin{equation}
\int f[g h k]_{\mathcal M_{\rm (I)}}(x)=-\int k[g h f]_{\mathcal M_{\rm (I)}}(x),
\end{equation}
because
\begin{equation}
f_{\rm (I)}\left(p,q,k\right)=-f_{\rm (I)}\left(p,q,-(p+q+k)\right).
\end{equation}
Consequently,
\begin{equation}
\begin{split}
S^{e^2}_{\rm (I)_r}=&-\frac{e^2}{4}\theta^{ij}\theta^{kl}\int\,
-\frac{1}{4}f^{\mu\nu}\left[f_{\mu\nu}f_{ik}f_{jl}\right]_{\star_{3'}}+\frac{1}{8}f^{\mu\nu}\left[f_{ij}f_{kl} f_{\mu\nu}\right]_{\star_{3'}}+\frac{1}{2}\theta^{pq}f^{\mu\nu}\left[\partial_i f_{jk} f_{lp}\partial_q f_{\mu\nu}\right]_{\mathcal M_{\rm (I)}}.
\end{split}
\end{equation}
Finally one can use \eqref{IID1} once more to show
\begin{equation}
\int\,\theta^{ij}\theta^{kl}f^{\mu\nu}\left[f_{ij}f_{kl} f_{\mu\nu}\right]_{\star_{3'}}=\frac{1}{2}\int\,\theta^{ij}\theta^{kl}f^{\mu\nu}(\left[f_{ij}f_{kl} f_{\mu\nu}\right]_{\star_{3'}}+\left[f_{kl}f_{ij} f_{\mu\nu}\right]_{\star_{3'}})=\int\,\theta^{ij}\theta^{kl}(f^{\mu\nu}\star_2 f_{ij})(f_{kl}\star_2 f_{\mu\nu}).
\end{equation}
Thus, in the end we indeed obtain \eqref{2.10} from \eqref{2.8}. 

Explicit gauge invariance of the action \eqref{2.9} can be verified either by adding the ambiguity terms derived in~\cite{Trampetic:2015zma} to \eqref{2.10}, or by repeating the procedure above and obtaining \eqref{2.11}, where a different 3-product $[fgh]_{\mathcal M_{\rm (II)}}(x)$ is defined as below
\begin{gather}
[f g h]_{\mathcal M_{\rm (II)}}(x)=\int\,e^{-i(p+q+k)x}\tilde f(p)\tilde g(q)\tilde h(k)f_{\rm (II)}\left(p,q,k\right),
\\
\begin{split}
f_{\rm (II)}\left(p,q,k\right)&=\frac{\sin\left(\frac{p\theta q}{2}-\frac{q\theta k}{2}\right)\sin\frac{p\theta k}{2}}{\frac{p\theta q}{2}\frac{p\theta k}{2}\left(\frac{p\theta q}{2}-\frac{q\theta k}{2}\right)\left(\frac{p\theta k}{2}+\frac{q\theta k}{2}\right)}-\frac{\sin\left(\frac{p\theta q}{2}+\frac{p\theta k}{2}\right)\sin\frac{q\theta k}{2}}{\frac{p\theta q}{2}\frac{q\theta k}{2}\left(\frac{p\theta q}{2}+\frac{p\theta k}{2}\right)\left(\frac{p\theta k}{2}+\frac{q\theta k}{2}\right)}.
\end{split}
\end{gather}
(Un)like the $\mathcal M_{\rm (I)}$ product, the $\mathcal M_{\rm (II)}$ product satisfies a cyclicity condition
\begin{equation}
\int f[g h k]_{\mathcal M_{\rm (II)}}(x)=\int k[g h f]_{\mathcal M_{\rm (II)}}(x),
\end{equation}
since
\begin{equation}
f_{\rm (II)}\left(p,q,k\right)=f_{\rm (II)}\left(p,q,-(p+q+k)\right).
\end{equation}

\section{VERTICES}
In this section we present the detailed definition of each terms in the four-photon
interaction vertex in the momentum space $\Gamma^{\mu_1\mu_2\mu_3\mu_4}_{\rm (I,II)}\left(p_1,p_2,p_3,p_4\right)$:
\begin{gather}
\begin{split}
\Gamma_A^{\mu_1\mu_2\mu_3\mu_4}\left(p_1,p_2,p_3,p_4\right)=f_{\star_2}\left(p_1,p_2\right)f_{\star_2}\left(p_3,p_4\right)
V_A^{\mu_1\mu_2\mu_3\mu_4}\left(p_1,p_2,p_3,p_4\right),
\\
\Gamma_{B}^{\mu_1\mu_2\mu_3\mu_4}\left(p_1,p_2,p_3,p_4\right)=f_{\star_2}\left(p_1,p_2\right)f_{\star_2}\left(p_3,p_4\right)
V_B^{\mu_1\mu_2\mu_3\mu_4}\left(p_1,p_2,p_3,p_4\right),
\end{split}
\label{A1}
\end{gather}
\begin{gather}
\begin{split}
\Gamma_1^{\mu_1\mu_2\mu_3\mu_4}\left(p_1,p_2,p_3,p_4\right)=&2f_{\star_{3'}}\left(p_2,p_3,p_4\right)V_1^{\mu_1\mu_2\mu_3\mu_4}\left(p_1,p_2,p_3,p_4\right),
\\
\Gamma_2^{\mu_1\mu_2\mu_3\mu_4}\left(p_1,p_2,p_3,p_4\right)=&(2f_{\star_2}\left(p_1,p_2\right)f_{\star_2}\left(p_3,p_4\right)-f_{\star_{3'}}\left(p_2,p_3,p_4\right)
-f_{\star_{3'}}\left(p_4,p_3,p_2\right))V_2^{\mu_1\mu_2\mu_3\mu_4}\left(p_1,p_2,p_3,p_4\right),
\\
\Gamma_3^{\mu_1\mu_2\mu_3\mu_4}\left(p_1,p_2,p_3,p_4\right)=&f_{\star_{3'}}(p_4,p_2,p_3)V_3^{\mu_1\mu_2\mu_3\mu_4}\left(p_1,p_2,p_3,p_4\right),
\\
\Gamma_4^{\mu_1\mu_2\mu_3\mu_4}\left(p_1,p_2,p_3,p_4\right)=&f_{\star_2}(p_1,p_2)f_{\star_2}(p_3,p_4)V_4^{\mu_1\mu_2\mu_3\mu_4}\left(p_1,p_2,p_3,p_4\right),
\\
\Gamma_5^{\mu_1\mu_2\mu_3\mu_4}\left(p_1,p_2,p_3,p_4\right)=&f_{\rm (I)}(p_2,p_3,p_4)V_5^{\mu_1\mu_2\mu_3\mu_4}\left(p_1,p_2,p_3,p_4\right),
\end{split}
\label{A2}
\end{gather}
\begin{gather}
\begin{split}
{\Gamma'}_1^{\mu_1\mu_2\mu_3\mu_4}\left(p_1,p_2,p_3,p_4\right)=&(4f_{\star_2}\left(p_1,p_2\right)f_{\star_2}\left(p_3,p_4\right)
-2f_{\star_{3}}\left(p_2,p_3,p_4\right))V_1^{\mu_1\mu_2\mu_3\mu_4}\left(p_1,p_2,p_3,p_4\right),
\\
{\Gamma'}_2^{\mu_1\mu_2\mu_3\mu_4}\left(p_1,p_2,p_3,p_4\right)=&2(f_{\star_{3}}\left(p_2,p_3,p_4\right) -f_{\star_2}\left(p_2,p_3\right)f_{\star_2}\left(p_1,p_4\right))V_2^{\mu_1\mu_2\mu_3\mu_4}\left(p_1,p_2,p_3,p_4\right),
\\
{\Gamma'}_3^{\mu_1\mu_2\mu_3\mu_4}\left(p_1,p_2,p_3,p_4\right)=&(3f_{\star_2}(p_1,p_2)-2f_{\star_3}(p_2,p_3,p_4))V_3^{\mu_1\mu_2\mu_3\mu_4}\left(p_1,p_2,p_3,p_4\right),
\\
{\Gamma'}_4^{\mu_1\mu_2\mu_3\mu_4}\left(p_1,p_2,p_3,p_4\right)=&(2f_{\star_2}(p_1,p_2)-f_{\star_3}(p_2,p_3,p_4))V_4^{\mu_1\mu_2\mu_3\mu_4}\left(p_1,p_2,p_3,p_4\right),
\\
{\Gamma'}_5^{\mu_1\mu_2\mu_3\mu_4}\left(p_1,p_2,p_3,p_4\right)=&f_{\rm (II)}(p_2,p_3,p_4){V'}_5^{\mu_1\mu_2\mu_3\mu_4}\left(p_1,p_2,p_3,p_4\right).
\end{split}
\label{A3}
\end{gather}
The $V_i$ tensor structures are listed next
\begin{gather}
\begin{split}
V_A^{\mu_1\mu_2\mu_3\mu_4}\left(p_1,p_2,p_3,p_4\right)=&(p_1 p_3)(p_2p_4)\theta^{\mu_1\mu_2}\theta^{\mu_3\mu_4}-(p_1 p_3)\theta^{\mu_1\mu_2}(\theta p_4)^{\mu_3}p_2^{\mu_4}
+(p_2 p_4)\theta^{\mu_1\mu_2}p_1^{\mu_3}(\theta p_3)^{\mu_4}
\\&+(p_3\theta p_4)\theta^{\mu_1\mu_2}p_1^{\mu_3}p_2^{\mu_4}
-(p_1 p_3)(\theta p_2)^{\mu_1}p_4^{\mu_2}\theta^{\mu_3\mu_4}+(p_1 p_3)(\theta p_2)^{\mu_1}(\theta p_4)^{\mu_3}g^{\mu_2\mu_4}
\\&-(\theta p_2)^{\mu_1}p_4^{\mu_2}p_1^{\mu_3}(\theta p_3)^{\mu_4}-(p_3\theta p_4)(\theta p_2)^{\mu_1}p_1^{\mu_3}g^{\mu_2\mu_4}
+(p_2 p_4)p_3^{\mu_1}(\theta p_1)^{\mu_2}\theta^{\mu_3\mu_4}
\\&-p_3^{\mu_1}(\theta p_1)^{\mu_2}(\theta p_4)^{\mu_3}g^{\mu_2\mu_4}
+(p_2 p_4)g^{\mu_1\mu_3}(\theta p_1)^{\mu_2}(\theta p_3)^{\mu_4}+(p_3\theta p_4)g^{\mu_1\mu_3}(\theta p_1)^{\mu_2}p_2^{\mu_4}
\\&+(p_1\theta p_2)p_3^{\mu_1}p_4^{\mu_2}\theta^{\mu_3\mu_4}-(p_1\theta p_2)p_3^{\mu_1}(\theta p_4)^{\mu_3}g^{\mu_2\mu_4}
+(p_1\theta p_2)g^{\mu_1\mu_3}p_4^{\mu_2}(\theta p_3)^{\mu_4}
\\&+(p_1\theta p_2)(p_3\theta p_4)g^{\mu_1\mu_3}g^{\mu_2\mu_4},
\end{split}
\label{A4}
\end{gather}
\begin{gather}
\begin{split}
V_{B}^{\mu_1\mu_2\mu_3\mu_4}\left(p_1,p_2,p_3,p_4\right)=&2(\theta p_1)^{\mu_1}((p_2 p_3)p_4^{\mu_2}\theta^{\mu_3\mu_4}-(p_2 p_3)(\theta p_4)^{\mu_3}g^{\mu_2\mu_4}+p_4^{\mu_2}p_2^{\mu_3}(\theta p_3)^{\mu_4}
\\&+(p_3\theta p_4)p_2^{\mu_3}g^{\mu_2\mu_4}-(p_2 p_4)p_3^{\mu_2}\theta^{\mu_3\mu_4}+p_3^{\mu_2}(\theta p_4)^{\mu_3}p_2^{\mu_4}-(p_2 p_4)g^{\mu_2\mu_3}(\theta p_3)^{\mu_4}
\\&-(p_3\theta p_4)g^{\mu_2\mu_3}p_2^{\mu_4}),
\end{split}
\label{A5}
\end{gather}
\begin{gather}
\begin{split}
V_1^{\mu_1\mu_2\mu_3\mu_4}\left(p_1,p_2,p_3,p_4\right)=2(p_1 p_2)&(p_3^{\mu_1}(\theta p_4)^{\mu_2}\theta^{\mu_3\mu_4}-p_3^{\mu_1}(\theta p_4)^{\mu_3}\theta^{\mu_2\mu_4}
+g^{\mu_1\mu_3}(\theta p_4)^{\mu_2}(\theta p_3)^{\mu_4}+(p_3\theta p_4)g^{\mu_1\mu_3}\theta^{\mu_2\mu_4})
\\-2p_1^{\mu_2}((&p_2\theta p_4)p_3^{\mu_1}\theta^{\mu_3\mu_4}+p_3^{\mu_1}(\theta p_4)^{\mu_3}(\theta p_2)^{\mu_4}
+(p_2\theta p_4)g^{\mu_1\mu_3}(\theta p_3)^{\mu_4}-(p_3\theta p_4)g^{\mu_1\mu_3}(\theta p_2)^{\mu_4})
\\+2(p_1 p_3&)(p_2^{\mu_1}(\theta p_4)^{\mu_3}\theta^{\mu_2\mu_4}-p_2^{\mu_1}(\theta p_4)^{\mu_2}\theta^{\mu_3\mu_4}
+g^{\mu_1\mu_2}(\theta p_4)^{\mu_3}(\theta p_2)^{\mu_4}+(p_2\theta p_4)g^{\mu_1\mu_2}\theta^{\mu_3\mu_4})
\\
-2p_1^{\mu_3}((&p_3\theta p_4)p_2^{\mu_1}\theta^{\mu_2\mu_4}+p_2^{\mu_1}(\theta p_4)^{\mu_2}(\theta p_3)^{\mu_4}
+(p_3\theta p_4)g^{\mu_1\mu_2}(\theta p_2)^{\mu_4}-(p_2\theta p_4)g^{\mu_1\mu_2}(\theta p_3)^{\mu_4}),
\\
V_2^{\mu_1\mu_2\mu_3\mu_4}\left(p_1,p_2,p_3,p_4\right)=(\theta p_4)^{\mu_3}&((p_1 p_2)p_4^{\mu_1}\theta^{\mu_2\mu_4}-(p_1 p_2)(\theta p_4)^{\mu_2}g^{\mu_1\mu_4}
+p_1^{\mu_2}(\theta p_2)^{\mu_4}p_4^{\mu_1}
+p_1^{\mu_2}(p_2\theta p_4)g^{\mu_1\mu_4}
\\
-(p_1 p_4)&p_2^{\mu_1}\theta^{\mu_1\mu_4}
+p_1^{\mu_4}p_2^{\mu_1}(\theta p_4)^{\mu_2}
-(p_1 p_4)(\theta p_2)^{\mu_4}g^{\mu_1\mu_2}-p_1^{\mu_4}(p_2\theta p_4)g^{\mu_1\mu_2}),
\\
V_3^{\mu_1\mu_2\mu_3\mu_4}\left(p_1,p_2,p_3,p_4\right)=-((p_1 p_4&)g^{\mu_1\mu_4}-p_1^{\mu_4}p_4^{\mu_1})((\theta p_2)^{\mu_3}(\theta p_3)^{\mu_2}+\theta^{\mu_2\mu_3}(p_2\theta p_3)),
\\
V_4^{\mu_1\mu_2\mu_3\mu_4}\left(p_1,p_2,p_3,p_4\right)=((p_1p_4)&g^{\mu_1\mu_4}-p_1^{\mu_4}p_4^{\mu_1})(\theta p_2)^{\mu_2}(\theta p_3)^{\mu_3},
\\
V_5^{\mu_1\mu_2\mu_3\mu_4}\left(p_1,p_2,p_3,p_4\right)=-((p_1 p_4&)g^{\mu_1\mu_4}-p_1^{\mu_4}p_4^{\mu_1})(\theta p_2)^{\mu_2}((\theta p_2)^{\mu_3}(p_3\theta p_4)+(\theta p_4)^{\mu_3}(p_2\theta p_3)),
\\
{V'}_5^{\mu_1\mu_2\mu_3\mu_4}\left(p_1,p_2,p_3,p_4\right)=((p_1 p_4&)g^{\mu_1\mu_4}-p_1^{\mu_4}p_4^{\mu_1})
((\theta p_2)^{\mu_2}(p_1\theta p_2)((\theta p_2)^{\mu_3}(p_3\theta p_4)+(\theta p_4)^{\mu_3}(p_2\theta p_3))
\\&\hspace{-1cm}-((\theta p_2)^{\mu_3}(p_3\theta p_4)+(\theta p_4)^{\mu_3}(p_2\theta p_3))((\theta p_3)^{\mu_2}(p_2\theta p_4)-(\theta p_4)^{\mu_2}(p_2\theta p_3))).
\end{split}
\label{A7}
\end{gather}

\section{Partial tensor reduction and integration results}

The partial $\tau(\mathcal T)$-tensor reduction results are listed below
\begin{equation}
\begin{split}
\tau_{\rm (I)}^{\mu\nu}=&-\frac{1}{2D}\{\left[g^{\mu\nu}p^2-p^{\mu}p^{\nu}\right](\tr\theta\theta)(\kappa_3-1)+[g^{\mu\nu}(\theta p)^2-(\theta\theta)^{\mu\nu}p^2
+ p^{\{\mu}(\theta\theta p)^{\nu\}}]4(\theta p)^2(\kappa_1-\kappa_2)
\\&+(\theta p)^{\mu}(\theta p)^{\nu}(4+(1-\kappa_3)D(D-1)-16\kappa+8\kappa^2+8(D-1)(\kappa_1-\kappa_2)+4\kappa_4)\},
\end{split}
\label{3.15}
\end{equation}
\begin{equation}
\begin{split}
\tau_{\rm (II)}^{\mu\nu}=&-\frac{1}{2D}\{\left[g^{\mu\nu}p^2-p^{\mu}p^{\nu}\right](\tr\theta\theta)3(\kappa_3-1)+[g^{\mu\nu}(\theta p)^2-(\theta\theta)^{\mu\nu}p^2
+ p^{\{\mu}(\theta\theta p)^{\nu\}}]8(\theta p)^2(\kappa'_1-\kappa'_2)
\\&+(\theta p)^{\mu}(\theta p)^{\nu}(3(1-\kappa'_3)D(D-1)-16\kappa+8\kappa^2+16(D-1)(\kappa'_1-\kappa'_2)+8\kappa'_4)\},
\end{split}
\label{3.16}
\end{equation}
\begin{equation}
\begin{split}
\mathcal T_{\rm (I)}^{\mu\nu\rho\sigma}=&-\frac{1}{2}\{(g^{\mu\nu}(\theta p)^{\rho}(\theta p)^{\sigma}+\theta^{\mu\rho}p^\nu(\theta p)^\sigma+\theta^{\nu\rho}p^\mu(\theta p)^\sigma+\theta^{\mu\rho}\theta^{\nu \sigma}p^2)2((D-3)\kappa^2-2\kappa+\kappa_1+\kappa_2)
\\&+(2g^{\mu\nu}p^{\rho}(\theta\theta p)^{\sigma}-g^{\mu\rho}p^{\nu}(\theta\theta p)^{\sigma}-g^{\nu\rho}p^{\mu}(\theta\theta p)^{\sigma}
-p^\mu(\theta\theta)^{\nu\rho}p^\sigma-p^\nu(\theta\theta)^{\mu\rho}p^\sigma
\\&+g^{\mu\rho}(\theta\theta)^{\nu\sigma}p^2+g^{\nu\rho}(\theta\theta)^{\mu\sigma}p^2)\cdot(2\kappa-\kappa_1-\kappa_2)
\\&+(g^{\mu\rho}(\theta p)^\nu(\theta p)^\sigma+g^{\nu\rho}(\theta p)^\mu(\theta p)^\sigma+\theta^{\mu\rho}(\theta p)^\nu p^\sigma+\theta^{\nu\rho}(\theta p)^\mu p^\sigma)
(-1-2\kappa_3+\kappa_4+(2+D)\kappa_1
\\&+(D-2)(\kappa_2-2\kappa)+4\kappa^2)
\\&+(g^{\mu\nu}g^{\rho\sigma}(\theta p)^2+(\theta\theta)^{\mu\nu}(p^\rho p^\sigma-p^2 g^{\rho\sigma})+(p^\mu(\theta\theta p)^\nu+p^\nu(\theta\theta p)^\mu)g^{\rho\sigma}
-g^{\mu\rho}g^{\nu\sigma}(\theta p)^2-g^{\mu\rho}(\theta\theta p)^\nu p^\sigma
\\&-g^{\nu\rho}(\theta\theta p)^\mu p^\sigma)\cdot 2\kappa^2
\\&+(\theta p)^\mu(\theta p)^\nu g^{\rho\sigma}\cdot 4(\kappa_1+\kappa_2-2\kappa+(D-1)\kappa_4)-\left[g^{\mu\nu}p^2-p^{\mu}p^{\nu}\right](\theta\theta)^{\rho\sigma}(\kappa_4-1)\},
\end{split}
\label{3.17}
\end{equation}
\begin{equation}
\begin{split}
\mathcal T_{\rm (II)}^{\mu\nu\rho\sigma}=&-\frac{1}{2}\{(g^{\mu\nu}(\theta p)^{\rho}(\theta p)^{\sigma}+\theta^{\mu\rho}p^\nu(\theta p)^\sigma+\theta^{\nu\rho}p^\mu(\theta p)^\sigma-\theta^{\mu\rho}\theta^{\nu\sigma}p^2)\cdot 2((D-3)\kappa^2-2\kappa+2\kappa'_2)
\\&+(2g^{\mu\nu}p^{\rho}(\theta\theta p)^{\sigma}-g^{\mu\rho}p^{\nu}(\theta\theta p)^{\sigma}-g^{\nu\rho}p^{\mu}(\theta\theta p)^{\sigma}
\\&-p^\mu(\theta\theta)^{\nu\rho}p^\sigma-p^\nu(\theta\theta)^{\mu\rho}p^\sigma
+g^{\mu\rho}(\theta\theta)^{\nu\sigma}p^2+g^{\nu\rho}(\theta\theta)^{\mu\sigma}p^2)\cdot2\left(\kappa-\kappa'_2\right)
\\&+(g^{\mu\rho}(\theta p)^\nu(\theta p)^\sigma+g^{\nu\rho}(\theta p)^\mu(\theta p)^\sigma+\theta^{\mu\rho}(\theta p)^\nu p^\sigma+\theta^{\nu\rho}(\theta p)^\mu p^\sigma)
\cdot(-2\kappa'_3-2(D-2)\kappa+4\kappa^2+2(D-2)\kappa'_2)
\\&+(g^{\mu\nu}g^{\rho\sigma}(\theta p)^2+(\theta\theta)^{\mu\nu}(p^\rho p^\sigma-p^2 g^{\rho\sigma})+(p^\mu(\theta\theta p)^\nu+p^\nu(\theta\theta p)^\mu)g^{\rho\sigma}
\\&-g^{\mu\rho}g^{\nu\sigma}(\theta p)^2-g^{\mu\rho}(\theta\theta p)^\nu p^\sigma-g^{\nu\rho}(\theta\theta p)^\mu p^\sigma)\cdot 2\kappa^2
\\&+(\theta p)^\mu(\theta p)^\nu g^{\rho\sigma}\cdot (8\kappa'_2-8\kappa+(D-1)(2\kappa'_3+\kappa'_4-3))-\left[g^{\mu\nu}p^2-p^{\mu}p^{\nu}\right](\theta\theta)^{\rho\sigma}(2\kappa'_3+\kappa'_4-3)\}.
\end{split}
\label{3.18}
\end{equation}

\end{document}